\documentclass[11pt]{article}
\usepackage{amsmath,amssymb,color,graphics,epsfig}
\usepackage{graphicx}
\usepackage{subfigure}

\usepackage{amsfonts}

\newcommand{\be}{\begin{equation}}
\newcommand{\ee}{\end{equation}}
\newcommand{\bea}{\setlength\arraycolsep{2pt} \begin{eqnarray}}
\newcommand{\eea}{\end{eqnarray}}
\newcommand{\nn}{\nonumber}

\def\ft#1#2{{\textstyle{\frac{\scriptstyle #1}{\scriptstyle #2} } }}
\def\fft#1#2{{\frac{#1}{#2}}}

\def\0{{\sst{(0)}}}
\def\1{{\sst{(1)}}}
\def\2{{\sst{(2)}}}
\def\3{{\sst{(3)}}}
\def\4{{\sst{(4)}}}
\def\5{{\sst{(5)}}}
\def\6{{\sst{(6)}}}
\def\7{{\sst{(7)}}}
\def\8{{\sst{(8)}}}
\def\sst#1{{\scriptscriptstyle #1}}

\newcommand{\eq}[1]{(\ref{#1})}

\newcommand{\bra}[1]{|#1\rangle}

\def\be{\begin{equation}}
\def\ee{\end{equation}}
\def\ba{\begin{eqnarray}}
\def\ea{\end{eqnarray}}

\def\nn{\nonumber}
\def\lf{\left}
\def\rt{\right}

\def\lf{\left}\def\rt{\right}\def\q{\theta}
\def\w{\omega}  \def\y {\psi}   \def\p {\pi} \def\a {\alpha}  \def\d {\delta} \def\f {\phi} \def\g {\gamma} \def\h
{\eta}   \def\l {\lambda}  \def\x {\xi} \def\c {\chi} \def\b {\beta} \def\n {\nu}
\def\m {\mu} \def\pd {\partial}   
 \def\W{\Omega} \def\Y {\Psi} 
   \def\D {\Delta}     

\def\.{\cdot}
\def\math {\mathcal}
\def\vp{\varphi}
\def\c{\mathcal{C}}

\begin{document}

\begin{flushright}
\end{flushright}

\vspace{25pt}
\begin{center}
{\large {\bf Circuit complexity for generalised coherent states in thermal field dynamics}}

\vspace{10pt}
Minyong Guo$^{1}$,  Zhong-Ying Fan$^2$, Jie Jiang$^{3}$, Xiangjing Liu$^{3}$, and Bin Chen$^{1,4,5}$\\

\vspace{10pt}
$^1${ Center for High Energy Physics, Peking University,\\
Beijing 100871, P. R. China\\}
$^2${ Center for Astrophysics, School of Physics and Electronic Engineering, \\
 Guangzhou University, Guangzhou 510006, China }\\
$^3${ Department of Physics, Beijing Normal University, \\
 Beijing 100875,  P. R. China}\\
$^4${Department of Physics and State Key Laboratory of Nuclear Physics and Technology, Peking University,\\
Beijing 100871, P. R. China}\\
$^5${Collaborative Innovation Center of Quantum Matter,\\
Beijing 100871, P. R. China}
\smallskip

\vspace{30pt}

\underline{ABSTRACT}
\end{center}

In this work, we study the circuit complexity for generalized coherent states in thermal systems by adopting the covariance matrix approach. We focus on the coherent thermal (CT) state, which is non-Gaussian and has a nonvanishing one-point function. We find that even though the CT state cannot be fully determined by the symmetric two-point function, the circuit complexity can still be computed in the framework of the covariance matrix formalism by properly enlarging the covariance matrix. Now the group generated by the unitary is the semiproduct of translation and the symplectic group. If the reference state is Gaussian, the optimal geodesic is still be generated by a horizontal generator such that the circuit complexity can be read from  the generalized covariance matrix associated to the target state by taking the cost function to be $F_2$.  For a single harmonic oscillator, we discuss carefully the complexity and its formation in the cases that the reference states are Gaussian and the target space is excited by a single mode or double modes. We show that the study can be extended to the free scalar field theory.

\rule[-8pt]{14.3cm}{0.05em}

Email: minyongguo@pku.edu.cn,\,\,fanzhy@gzhu.edu.cn,\,\,jiejiang@mail.bnu.edu.cn,\,\,\\$~~~~~~~~~~~~~~~~$liuxj@mail.bnu.edu.cn,\,\,bchen01@pku.edu.cn.\\$~~~~$
\rule[+12pt]{14.5cm}{0.05em}

\thispagestyle{empty}

\tableofcontents
\addtocontents{toc}{\protect\setcounter{tocdepth}{2}}




\section{Introduction}


Complexity has been a focus in the recent study of the AdS/CFT correspondence \cite{Maldacena:1997re} and blackhole physics. As firstly pointed out by L. Susskind \cite{Susskind:2014rva},  ``entanglement is not enough" to describe the dynamics of the blackhole, especially the growth of the Einstein-Rosen-Bridge(ERB). Instead, he proposed \cite{Susskind:2014moa} that the growth of the ERB should be dual to the growth of the quantum complexity of the evolving state, the thermofield double (TFD) state \cite{Maldacena:2001kr}.  There are two proposals put forward by Susskind and his collaborators to quantify the size of the ERB: one is the ``complexity=volume"(CV) conjecture \cite{Stanford:2014jda} which states that the holographic complexity is given by the volume of the codimension-1 maximal spacelike surface in the bulk connecting the left and right sides; the other is the ``complexity=action"(CA) conjecture \cite{Brown:2015bva,Brown:2015lvg} which states that  the holographic complexity is captured by the gravitational action of the bulk region known as the Wheeler-de Witt (WdW) patch bounded by light sheets. Both conjectures introduce the gravitational observables which probe the spacetime region deep behind the black hole horizon, and therefore they have been intensely discussed since their birth \cite{Fan:2019aoj,Fan:2018wnv,Fan:2018xwf,Moosa:2017yiz,HosseiniMansoori:2017tsm,Mahapatra:2018gig,Chapman:2016hwi,Carmi:2016wjl,Yang:2016awy,
Moosa:2017yvt,Swingle:2017zcd,Alishahiha:2018tep,An:2018xhv,Jiang:2018pfk,Yang:2019gce,Guo:2019vni,Cai:2016xho,
Lehner:2016vdi,Huang:2016fks,Cano:2018aqi,Jiang:2018sqj,Jiang:2019fpz,Feng:2018sqm,Alishahiha:2017hwg,Carmi:2017jqz}. Nevertheless, the development is hindered by our poor understanding the quantum complexity in the dual field theory.

Originally the complexity is a concept in theoretical computer science \cite{Arora2009,Moore2011}, characterizing the difficulty in carrying out a task. In quantum computing, one may find a unitary operation $\hat{U}$ which maps an input quantum state for some number of qubits to an output quantum state with the same number of qubits \cite{Aaronson:2016vto,Watrous2009,Dean2016}. In a circuit model, $\hat{U}$ could be constructed from some elementary gates. There could be many ways in constructing $\hat{U}$ to some accuracy $\epsilon>0$. The circuit complexity of the unitary $\hat{U}$ is given by the minimal number of elementary gates required to construct the desired unitary $\hat{U}$, up to some tolerance $\epsilon$. However, to generalize the above definition to  quantum field theory, even a free field theory, is highly nontrivial, due to the fact that there are infinite number of degrees of freedom in a field theory. In order to define the circuit complexity, one first needs to identify the reference state and the target state,  and then identify the optimal circuit out of the infinite number of possible circuits connecting the reference state and the final target state.

There have been some initial steps in studying the complexity in quantum field theory. In \cite{Jefferson:2017sdb}, the circuit complexity of the ground state of a free scalar field theory was investigated. The optimal circuit was determined geometrically by the minimal geodesic in the space of unitaries $\hat{U}$ with a suitable metric, as developed by Nielsen and his collaborators \cite{Nielsen}. This approach has been applied to free fermionic theories in \cite{Hackl:2018ptj,Khan:2018rzm}. Another similar geometric definition of the complexity\footnote{There is a complementary approach to understand the complexity in quantum field theory using path-integral techniques, see \cite{Caputa:2017yrh, Czech:2017ryf,Bhattacharyya:2018wym}. In addtion, the authors  developed a framework that enabled a definition of complexity for strongly coupled large N systems, i.e. holographic CFTs in \cite{Belina:2018, Belina:20182}. } based on the Fubini-Study metric has been explored in the free scalar field theory in \cite{Chapman:2017rqy}. The circuit complexity in interacting field theories has been discussed in \cite{Bhattacharyya:2018bbv}.

The study of the complexity  has been generalized to the TFD state in free scalar field theory \cite{Chapman:2018hou,Yang:2017nfn,Kim:2017qrq,Doro:2019}. In this case, the target state is the TFD state, while the reference state has different choices: in \cite{Chapman:2018hou} the reference state was chosen to be composed of two copies of the reference state used in \cite{Jefferson:2017sdb,Chapman:2017rqy}; in \cite{Yang:2017nfn,Kim:2017qrq} the reference state was two unentangled copies of the vacuum state. Due to the difference in reference state and other points, the complexity for the TFD state in two approaches differs in many ways, especially the one for the time-dependent TFD state.

 In this work, we would like to study the circuit complexity of general coherent state, extending the study of complexity of coherent state in \cite{Guo:2018kzl}. We will focus on the coherent states in thermal systems. There are two kinds of coherent states in a thermal system: the coherent thermal (CT) state and thermal coherent state. As these two kinds of states are somehow equivalent, we will consider the circuit complexity for coherent thermal state by applying the covariance matrix approach developed in \cite{Chapman:2018hou}. Since the one-point function of the CT state is not vanishing, the two-point function is not enough to characterize the state. Nevertheless we show that the covariance matrix formalism is still applicable after some improvement. The essential point is that if the reference state remains Gaussian, the optimal geodesic will still be generated by a horizontal generator such that the circuit complexity can be read from the norm of the generator.

 For a simple harmonic oscillator system, we compute the complexity  and discuss the formation of the complexity by choosing various Gaussian reference states and the target states with single mode and double modes. We extend our study on the complexity of the CT state to the free scalar field theory by fixing the reference state to be the Dirac vacuum state.

 The remaining parts of the paper are organized as follows. In section 2, we give a brief introduction to the coherent state and the thermal vacuum state in the harmonic oscillator system. In section 3, we introduce the general coherent states in thermal systems. In section 4, we study the circuit complexity for the coherent thermal state. Due to the loss of Gaussianity of the CT state, we need to generalize the covariance matrix approach and furthermore compute the circuit complexities for the Gaussian reference states. In section 5, we study the complexity of CT state in a free scalar field theory by choosing the Dirac vacuum state to be the reference state. We end with conclusions and discussions in section 6.

\section{Preliminaries: coherent state and thermal vacuum state}

 In this section, we shall briefly review the construction of the Glauber coherent state and the thermal vacuum state. It will be shown in the next section that proper considerations from these two states lead to several different generalizations of the coherent state to the thermal field dynamics. Besides, to study the circuit complexity of the coherent thermal state for a free field theory, we will begin our story with a toy model: the harmonic oscillators. The complexity for the field theory will be discussed at last in Sec. \ref{secqft}.

\subsection{Coherent state}
For a single harmonic oscillator with a Hamilton $H=p^2/2m+\ft 12 m \omega^2 q^2$, the annihilation and creation operator can  be defined by
\bea\label{aadag} &&a\equiv \fft{1}{\sqrt{2m\omega}}\big(m\omega q+i p \big)\,,\nn\\
&&a^\dag \equiv \fft{1}{\sqrt{2m\omega}}\big(m\omega q-i p \big)  ,\eea
with $p=-i\partial_q$.
They satisfy the commutation relation
\be [a\,,a^\dag]=1\,.\ee
The Hamilton can be expressed into the form of
\be H=\omega\big(a^\dag a+\ft 12 \big)   \,.\ee
The vacuum state is defined by
\be a|0\rangle =0 \,.\ee
The energy eigenstates are defined by the creation operators acting on the vacuum
\be\label{nstate} |n\rangle=\fft{1}{\sqrt{n!}}\big(a^\dag \big)^n |0\rangle \,.\ee
It is known that these states form a complete basis in the Hilbert space, namely
\be\label{basis} \sum_{n=0}^{\infty} |n\rangle\langle n|=1  \,.\ee

The Glauber coherent state is another kind of interesting excited state. It is defined by the eigenstate of the annihilation operator
\be a|\alpha\rangle\equiv \alpha|\alpha\rangle \,,\label{coherent0}\ee
 where in general $\alpha$ is a complex $c$-number since the operator $a$ is not Hermitian. In fact, the coherent state can also be obtained by a particular operator $D(\alpha)$ acting on the vacuum state
 \be |\alpha\rangle=D(\alpha)|0\rangle \label{coherent1}\,,\ee
where $D(\alpha)$ is called {\it displacement}, defined as
\be
D(\alpha)\equiv \mathrm{exp}(\alpha a^\dag-\alpha^* a).
\ee
 Note that $D(\alpha)$ is anti-Hermitian because of
$D^\dag(\alpha)=D^{-1}(\alpha)$ and it obeys
\be D(\alpha)a^{(\dag)} D^{-1}(\alpha)=a^{(\dag)}-\alpha^{(*)}\,.\ee
By simple calculations, one finds that the coherent state is a superposition of the energy eigenstates
\be\label{coherent}|\alpha\rangle=D(\alpha)|0\rangle=e^{-|\alpha|^2/2} e^{\alpha a^\dag}|0\rangle= e^{-|\alpha|^2/2}\sum_{n=0}^\infty \fft{\alpha^n}{\sqrt{n!}}|n\rangle \,,\ee
where in the second equality, we have adopted the relation $e^{A+B}=e^A e^B e^{-[A\,,B]/2}$, which holds when the commutator $[A\,,B]$ commutes with both $A$ and $B$. Using these results, it is easy to show that $(\ref{coherent1})$ is equivalent to (\ref{coherent0}) and hence can be viewed as an alternative definition for the coherent state. In fact, in our opinion, (\ref{coherent1}) might be a better one since it is more enlightening for generalisations to thermal field dynamics.
In addition, the time-dependent coherent state can be obtained as
\bea\label{timecoherent}
|\alpha(t)\rangle&=&e^{-i H t}|\alpha\rangle \nn\\
&=&e^{-i\omega t/2} e^{-|\alpha|^2/2}\sum_{n=0}^\infty \fft{\Big( \alpha e^{-i \omega t} \Big)^n }{\sqrt{n!}}|n\rangle\nn\\
&=&e^{-i\omega t/2} | \alpha e^{-i\omega t} \rangle\,.
\eea

 To end this subsection, we would like to introduce a new set of states $\{ |n\,,\alpha\rangle \}$ that is of great importance in the construction of generalised coherent states in thermal field dynamics. The states are produced by the displacement operator acting upon the energy eigenstates
\be\label{ncoherent1} |n\,,\alpha\rangle \equiv D(\alpha)|n\rangle \,,\ee
where $n=0\,,1\,,2\,,\cdots$. These states are complete as well
\be \sum_{n=0}^\infty|n\,,\alpha\rangle \langle n\,,\alpha|=1 \,.\ee
In fact, this set of states can be created and annihilated by the following operators
\bea
b^{(\dag)}&\equiv& D(\alpha)a^{(\dag)} D^{-1}(\alpha)=a^{(\dag)}-\alpha^{(*)} \, ,
\eea
In this case, the Glauber coherent state can be viewed as a ground state because of
\be b |\alpha\rangle =0 \,.\ee
Moreover, one has
\bea
&&b |n\,,\alpha\rangle=\sqrt{n}\,|n-1\,,\alpha\rangle\,,\nn\\
&&b^\dag |n\,,\alpha\rangle=\sqrt{n+1}\,|n+1\,,\alpha\rangle \,,
\eea
so that
\be\label{ncoherent2} |n\,,\alpha\rangle=\fft{\big(b^\dag \big)^n}{\sqrt{n!}}|\alpha\rangle \,.\ee
These relations are similar to those between the energy eigenstates and the operators $a\,,a^\dag$.

\subsection{Thermal vacuum state}

In thermal field dynamics, the finite temperature problems are treated by using the techniques developed for zero temperature quantum field theories. The price for this convenience is that one needs to deal with an enlarged Hilbert space, which is a direct product of two copies of the ordinary zero temperature Hilbert space. We denote
\be \mathcal{H}\equiv \mathcal{H}_L \otimes\mathcal{H}_R \,,\ee
where $\mathcal{H}_L$ and $\mathcal{H}_R$ stand for the ordinary Hilbert space for the zero temperature theory on the left hand side and the right hand side, respectively. In the following, all the operators and the state vectors for the left/right hand side will be assigned with a subscript ``$L/R$". The creation and annihilation operators obey the commutation relations
\be [a_L\,,a_L^\dag]=[a_R\,,a_R^\dag]=1\,,\quad [a_L\,,a_R^\dag]=[a_R\,,a_L^\dag]=0  \,.\ee
It is worth emphasizing that the thermal vacuum state is a kind of excited states, rather than the true vacuum state of the Hilbert space $\mathcal{H}_L\otimes \mathcal{H}_R$.
This will be clear from the relation (\ref{tfd1}). In recent literatures, it is usually called the Thermal Field Double (TFD) state. We will use this for a shorthand notation throughout this paper.

To build the TFD state, we first introduce an anti-Hermitian operator
\be U(\beta)=\mathrm{exp}\Big[\theta(\beta)\big(a_L^\dag a_R^\dag-a_L a_R \big) \Big] \,,\ee
where $\beta=1/T$ is the inverse of temperature and $\theta$ is related to the temperature by
\bea
&&\cosh\theta(\beta)=\Big( 1-e^{-\beta\omega} \Big)^{-1/2}\,,\nn\\
&&\sinh\theta(\beta)=\Big( e^{\beta\omega}-1 \Big)^{-1/2}\,.
\eea
Note that $\tanh{\theta}=e^{-\beta\omega/2}$. Under the Bogoliubov transformation, the creation and annihilation operators are transformed as\footnote{The relations can be derived by using the Baker-Campbell-Hausdorff(BCH) formula $e^A B e^{-A}=\sum_{i=0}^\infty \fft{1}{i!}[A^{(i)}\,,B]=B+\fft{1}{1!}[A,B]+\fft{1}{2!}[A,[A,B]]+\fft{1}{3!}[A,[A,[A,B]]]+\cdots$, .}
\bea\label{bogliubov}
&&a_L\rightarrow a_L(\beta)=U(\beta)a_L U(\beta)^\dag=\cosh\theta\, a_L-\sinh\theta\, a_R^\dag\,,\nn\\
&&a_R\rightarrow a_R(\beta)=U(\beta)a_R U(\beta)^\dag=\cosh\theta\, a_R-\sinh\theta\, a_L^\dag\,.
\eea
The new operators obey the following commutation relations
\be [a_L(\beta)\,,a_L^\dag(\beta)]=[a_R(\beta)\,,a_R^\dag(\beta)]=1\,,\quad [a_L(\beta)\,,a_R^\dag(\beta)]=[a_R(\beta)\,,a_L^\dag(\beta)]=0  \,.\ee
The TFD state is defined by
\be a_L(\beta)|\mathrm{TFD}\rangle=a_R(\beta)|\mathrm{TFD}\rangle=0 \,,\ee
or by the operator $U(\beta)$ acting on the vacua
\be\label{tfd1} |\mathrm{TFD}\rangle=U(\beta)|0\rangle_L|0\rangle_R \,.\ee
The two definitions are equivalent, as one can check using the Bogoliubov transformation. Furthermore, the TFD state can be written explicitly as
\bea\label{tfd2}
|\mathrm{TFD}\rangle&=&\mathrm{exp}\Big[\theta\big(a_L^\dag a_R^\dag-a_L a_R \big)\Big]|0\rangle_L|0\rangle_R \nn\\
&=&\big(\cosh\theta\big)^{-1}\,\mathrm{exp}\Big( \tanh\theta a_L^\dag a_R^\dag \Big)|0\rangle_L|0\rangle_R \nn\\
&=&\Big(1-e^{-\beta\omega}\Big)^{1/2}\sum_{n=0}^\infty e^{-n\beta\omega/2}|n\rangle_L|n\rangle_R\,,
\eea
where in the second equality we have adopted the operator identity
\bea\label{identity}
\lefteqn{\mathrm{exp}\Big[\theta\big(a_L^\dag a_R^\dag-a_L a_R \big)\Big]}\\
&=&\mathrm{exp}\Big(\tanh\theta a_L^\dag a_R^\dag \Big)
\mathrm{exp}\Big[ -\big(a_L^\dag a_L+ a_R^\dag a_R+1\big)\log\cosh\theta \Big]\mathrm{exp}\Big(-\tanh\theta a_L a_R \Big)\,.\nn
\eea
However, since the state is introduced to deal with the finite temperature problems, it is interesting to see how the state become for a local observer, for example on the left hand side. Tracing over the degrees of freedom on the right hand side, one finds the reduced density matrix
\bea
\rho_L&=&\mathrm{Tr}_{\mathcal{H}_R}|\mathrm{TFD}\rangle \langle \mathrm{TFD}| \nn\\
&=&\Big(1-e^{-\beta\omega}\Big)\sum_{n=0}^\infty e^{-n\beta\omega}(|n\rangle\langle n|)_L\,.
\eea
Clearly, this is a thermal density matrix, describing an ordinary thermal equilibrium state.
 In fact, this should be a priori for the construction of TFD state in thermal field dynamics. Likewise, it is a priori rule to test the generalisations of the coherent state: a suitable generalisation should not only be coherent in thermal field dynamics but also be thermal for the one-sided theory.

\subsubsection{Time dependent TFD state}

Since the theory under consideration is a direct product of two copies of ordinary zero temperature theory, the time evolution operator is given by $e^{-i(H_L t_L+H_R t_R)}$. The time dependent TFD state is obtained by
\be |\psi_{TFD}(t) \rangle=e^{-i(H_L t_L+H_R t_R)}|\mathrm{TFD}\rangle  \,.\ee
In principle, the evolution on one side is independent from the one on the other. In this paper, we choose the symmetric case $t_L=t_R=t/2$ for the sake of convenience. It was established \cite{Chapman:2018hou} that the time dependent TFD state can be written into a nice form similar to (\ref{tfd1}), by using the explicit expression (\ref{tfd2}) for the TFD state. However, in the following, we would like to provide a different derivation by using the operator algebra directly. The new approach is more neat and more suitable for our later purpose.

By setting $\bar H=(H_L+H_R)/2$, one finds
\bea\label{timetfd}
|\psi_{TFD}(t) \rangle&=&e^{-i\bar H t} |\mathrm{TFD}\rangle \nn\\
&=&e^{-i\bar H t}U(\beta)e^{i\bar H t}e^{-i\bar H t} |0\rangle_L |0\rangle_R  \nn\\
&\equiv &e^{-i\omega t/2} U(\beta\,,t) |0\rangle_L |0\rangle_R \,,
\eea
where $U(\beta\,,t)\equiv e^{-i\bar H t}U(\beta)e^{i\bar H t}$. After introducing the operators
\be B_\pm=a_L^\dag a_R^\dag\pm a_L a_R \,,\ee
which obey the commutation relation
\be [-i\bar H t\,,B_\pm]=-i\omega t B_\mp \,,\ee
 one obtains
\bea
U(\beta\,,t)&=&e^{-i\bar H t}U(\beta)e^{i\bar H t} \nn\\
&=&\mathrm{exp}\Big[\theta\, e^{-i\bar H t}B_- e^{i\bar H t}\Big]\nn\\
&=&\mathrm{exp}\big(z a_L^\dag a_R^\dag-z^* a_L a_R \big)\,,
\eea
where $z=\theta e^{-i\omega t}$ and in the last line we have adopted the Baker-Campbell-Hausdorff (BCH) formula to deduce the relation
\be  e^{-i\bar H t}B_- e^{i\bar H t}=e^{-i\omega t}a_L^\dag a_R^\dag- e^{i\omega t} a_L a_R \,.\ee
Moreover, it turns out that the operator $U(\beta\,,t)$ can be recast into a more compact Gaussian form in terms of the canonical variables $\xi^a=(q_L\,,q_R\,,p_L\,,p_R)$,
\be
U(\beta\,,t)=e^{-\fft{i}{2}k_{ab}^{(0)}\xi^a \xi^b},
\ee
  where
\bea\label{XT}
k_{ab}^{(0)}=\theta\left(
\begin{array}{cccc}
  0 & m\omega \sin(\omega t)&0 & \cos(\omega t) \\
  m\omega \sin(\omega t) &0 & \cos(\omega t)& 0 \\
  0 & \cos(\omega t)& 0 & -\ft{\sin(\omega t)}{m\omega} \\
  \cos(\omega t) & 0& -\ft{\sin(\omega t)}{m\omega} & 0
\end{array}
\right)\,.
\eea
In other words, the TFD state is a Gaussian state up to an unimportant phase factor because of $|\psi_{TFD}(t) \rangle=e^{-\fft{i}{2}k_{ab}^{(0)}\xi^a \xi^b}|0\rangle_L |0\rangle_R $. Expressing the state into this form is particularly useful when computing the complexity in the covariance matrix approach \cite{Chapman:2018hou}. We will turn to this point in Sec. \ref{cma}.

\section{Generalised coherent states in thermal field dynamics}

Now we are ready to introduce the generalised coherent states in thermal field dynamics. Interestingly, we find that
there are two types of generalisations in literature \cite{tcs1,tcs2,tcs3,tcs4,tcs5,tcs6,tcs7,tcs8}. In \cite{tcs8}, the generalized states are called Coherent Thermal (CT) state and Thermal Coherent (TC) state, respectively. However, we will show that though the two states are indeed defined differently, they can be related to each other via a parameter transformation, see (\ref{pptrans}) and (\ref{connection}). As a consequence,  the two definitions are equivalent in the sense that they just scan the eigenvalue spaces for the same set of states in different ways, leading to the apparent differences.

 Before introducing these states, we explain some of our notations below at first. The ordinary Glauber coherent state will be represented as $|\alpha\rangle_L |\gamma\rangle_R$, where $\alpha\,,\gamma$ characterize the eigenvalues of the left- and right-hand-side annihilation operators respectively. The state is obtained by acting the displacement on the vacua
 \be|\alpha\rangle_L |\gamma\rangle_R=D(\alpha\,,\gamma)|0\rangle_L |0\rangle_R\,,\ee
where $D(\alpha\,,\gamma)$ is the product of the displacements on two sides
\be D(\alpha\,,\gamma)\equiv D_L(\alpha)D_R(\gamma)=\mathrm{exp}\big[\alpha a_L^\dag+\gamma a_R^\dag-\alpha^* a_L-\gamma^* a_R \big] \,.\ee

\subsection{Coherent thermal state}
A CT state is defined by an anti-Hermitian operator $\widetilde{U}(\beta)$ acting upon the Glauber coherent state \cite{tcs8}
\be |\mathrm{CT}\rangle\equiv \widetilde{U}(\beta)|\alpha\rangle_L |\gamma\rangle_R \,,\quad \widetilde{U}(\beta)=\mathrm{exp}\Big[\theta(\beta)\big(b_L^\dag b_R^\dag-b_L b_R \big) \Big]\,,\ee
where it was understood that
\bea
&&b_L=D_L(\alpha)a_L D_L^\dag(\alpha)=a_L-\alpha\,,\nn\\
&&b_R=D_R(\gamma)a_R D_R^\dag(\gamma)=a_R-\gamma \,.
\eea
The operator $\widetilde{U}(\beta)$ is related to $U(\beta)$ via a unitary transformation
\be \widetilde{U}(\beta)=D(\alpha\,,\gamma)U(\beta)D^\dag(\alpha\,,\gamma) \,.\ee
Therefore, one has
\bea\label{CTS1}
|\mathrm{CT}\rangle&=&D(\alpha\,,\gamma)U(\beta)|0\rangle_L |0\rangle_R=D(\alpha\,,\gamma)|\mathrm{TFD}\rangle\,.
\eea
In other words, the state is first thermalized and then displaced. Moreover, by using Eq. (\ref{tfd2}) for the TFD state, the CT state can be expressed into a similar nice form
\be\label{CTS2} |\mathrm{CT}\rangle=\big(1-e^{-\beta\omega}\big)^{1/2}\sum_{n=0}^\infty e^{-n\beta\omega/2}|n\,,\alpha\rangle_L |n\,,\gamma\rangle_R\,. \ee
It is clear that the CT state is a two-parameter generalisation of the thermal vacuum state. On one hand,
when $\alpha=\gamma=0$, it reduces to the latter. On the other hand, the state for a local observer becomes thermal because of
\bea
\rho_L&=&\mathrm{Tr}_{\mathcal{H}_R}|\mathrm{CT}\rangle \langle \mathrm{CT}| \nn\\
&=&\Big(1-e^{-\beta\omega}\Big)\sum_{n=0}^\infty e^{-n\beta\omega}(|n\,,\alpha\rangle\langle n\,,\alpha|)_L\,.
\eea
Note that it describes the thermal equilibrium in terms of the set of states $\{|n\,,\alpha\rangle\}$.

The time-dependent coherent thermal state is produced by
\ba\begin{aligned}
|\psi_{CT}(t)\rangle&=e^{-i\bar H t}\, |\mathrm{CT}\rangle\\
&=e^{-i\bar H t} D(\a,\g)e^{i\bar H t}e^{-i\bar H t} |\mathrm{TFD}\rangle\\
&= \math{D}(\a,\g;\,t)|\psi_{TFD}(t) \rangle\,,
\end{aligned}\label{ctstime}\ea
where
\ba\begin{aligned}
\math{D}(\a,\g;\,t)\equiv e^{-i\bar H t} D(\a,\g)e^{i\bar H t}\,.
\end{aligned}\ea
Using the relations
\ba\begin{aligned}
e^{-i\bar H t} a_{L\,,R}\, e^{i\bar H t}=e^{i \w t/2}\, a_{L\,,R} \,,\ \ \ e^{-i\bar H t} a_{L\,,R}^\dag \,e^{i\bar H t}=e^{-i \w t/2}\,a_{L\,,R}^\dag \,,
\end{aligned}\label{aLR}\ea
we deduce
\ba\begin{aligned}
\math{D}(\a,\g;\,t)&=\mathrm{exp}\left[\alpha e^{-i\w t/2} a_L^\dag+\gamma e^{-i\w t/2} a_R^\dag-\alpha^* e^{i\w t/2} a_L-\gamma^*e^{i\w t/2} a_R \right]\\
\end{aligned}\,.\label{ctstimed}\ea
Furthermore, using Eq.\eq{aadag}, it can be expressed as an exponential of the superposition of canonical variables in the phase space $\xi^a=\{q_L\,,q_R\,,p_L\,,p_R\}$
\ba
\mathcal{D}(\alpha\,,\gamma;\,t)=e^{-i\l_a \x^a}=\exp\lf[-i\lf(\l_{q_L} q_L+\l_{q_R} q_R+\l_{p_L} p_L+\l_{p_R} p_R\rt)\rt]\,,
\ea
where
\ba\begin{aligned}
\l_{q_L}&=\sqrt{2m\w}\lf[\Re\a \sin\lf(\frac{\w t}{2}\rt)-\Im\a \cos\lf(\frac{\w t}{2}\rt)\rt]\,,\\
\l_{q_R}&=\sqrt{2m\w}\lf[\Re\g \sin\lf(\frac{\w t}{2}\rt)-\Im\g \cos\lf(\frac{\w t}{2}\rt)\rt]\,,\\
\l_{p_L}&=\sqrt{\frac{2}{m\w}}\lf[\Re\a \cos\lf(\frac{\w t}{2}\rt)+\Im\a \sin\lf(\frac{\w t}{2}\rt)\rt]\,,\\
\l_{p_R}&=\sqrt{\frac{2}{m\w}}\lf[\Re\g \cos\lf(\frac{\w t}{2}\rt)+\Im\g \sin\lf(\frac{\w t}{2}\rt)\rt]\,,\\
\end{aligned}\label{CTSlambda}\ea
where $\Re f$ and $\Im f$ denote the real and the imaginary part of $f$, respectively. From this, it is clear that the CT state is non-Gaussian. In sec. \ref{cma}, we will adopt the above result to compute the complexity of a CT state using the covariance matrix approach.

\subsection{Thermal coherent state}
Compared to the CT state, the thermal coherent (TC) state is defined by thermalizing a Glauber coherent state \cite{tcs8}
\be\label{tfdc} |\mathrm{TC}\rangle=U(\beta)\,|\alpha\rangle_L |\gamma\rangle_R=U(\beta)D(\alpha\,,\gamma)|0\rangle_L |0\rangle_R \,.\ee
Note the different order of the operators $U(\beta)\,, D(\alpha\,,\gamma)$ acting on the vacuum state in (\ref{CTS1}). According to the thermal Bogliubov transformation (\ref{bogliubov}), the TC state turns out to be the eigenstate of thermal annihilation operators, namely
\bea
&&a_L(\beta)|\mathrm{TC}\rangle=\alpha |\mathrm{TC}\rangle \,,\nn\\
&&a_R(\beta)|\mathrm{TC}\rangle=\gamma |\mathrm{TC}\rangle \,.
\eea
In addition, from (\ref{tfdc}), the TC state is related to the thermal vacuum state via a thermal displacement operator $\mathcal{D}(\alpha\,,\gamma;\,\beta)$
\be |\mathrm{TC}\rangle=\math{D}(\a,\g;\,\beta)|\mathrm{TFD}\rangle\,,\ee
where $\math{D}(\alpha\,,\gamma;\,\beta)$ is defined by
\ba\begin{aligned}
\math{D}(\a,\g;\,\beta)&\equiv U(\beta)D(\alpha\,,\gamma)U^\dag(\beta)\nn\\
&=\mathrm{exp}\Big[\alpha a_L^\dag(\beta)+\gamma a_R^\dag(\beta)-\alpha^* a_L(\beta)-\gamma^* a_R(\beta) \Big]\\
&=\mathrm{exp}\Big[\hat\alpha(\beta)  a_L^\dag+\hat\gamma(\beta)  a_R^\dag-\hat\alpha(\beta)^* a_L-\hat\gamma(\beta)^* a_R \Big]\,,\\
\end{aligned}\ea
with
\ba\begin{aligned}
\hat\alpha(\beta)&=\cosh\theta\, \a+\sinh\theta\, \g^*\,,\\
\hat\gamma(\beta)&=\cosh\theta\,\g+\sinh\theta\,\a^*\,.
\end{aligned}\label{pptrans}\ea
It is interesting to note that the thermal displacement $\math{D}(\a\,,\g;\,\b)$ is related to ordinary displacement $D(\alpha\,,\gamma)$ via the parameter transformation (\ref{pptrans}). One has
\be\math{D}(\a\,,\g;\,\b)=D(\hat\alpha\,,\hat\gamma \big)\,.\ee
Therefore, a TC state is related to a CT state by
\be\label{connection} |\mathrm{TC}(\alpha\,,\gamma )\rangle =|\mathrm{CT}(\hat\alpha\,,\hat\gamma)\rangle \,.\ee
 It strongly implies that the two set of states are equivalent. Of course, a particular TC state with given parameters $\alpha\,,\gamma$ is distinguished from the CT state with the same parameters since their dependence on the temperature are very different. However, the relation (\ref{connection}) tells us that the two sets of states just scan the eigenvalue space for the same set of states in different ways, leading to the apparent differences.

 In fact, the equivalence between the two sets of states becomes even more clear for several special cases. For example, when $\gamma=\pm\alpha^*$, it was argued \cite{tcs8} that both the TC state and the CT state belong to the generalized coherent states with respect to the Lie group $\mathrm{E}(1\,,1)$. We define
\be J_\theta=b_L^\dag b_R^\dag-b_L b_R\,,\quad J_\pm=b_L^\dag\mp b_R\,,\quad J_\pm^\dag=b_L\mp b_R^\dag \,.\ee
It follows that the generators $\{J_\theta\,,J_+\,,J_+^\dag \}$ or $\{J_\theta\,,J_-\,,J_-^\dag \}$ generate the Lie algebra of $\mathrm{E}(1\,,1)$
\bea
&&[J_\theta\,,J_+^{(\dag)}]=J_+^{(\dag)}\,,\quad [J_+\,,J_+^\dag]=0\,,\nn\\
&&[J_\theta\,,J_-^{(\dag)}]=J_-^{(\dag)}\,,\quad [J_-\,,J_-^\dag]=0\,.
\eea
Thus, in this case, the two states can be collectively represented as
\be \mathrm{exp}\Big(\sum_{i=1}^3 \epsilon_i J_i \Big) \,.\ee
From this, it is obvious that the two states must be related via a transformation of parameters since they both give a representation of the same group. Another interesting case is when $\alpha$, $\gamma$ are real. The parameter transformation (\ref{pptrans}) reduces to a boost. The Lie algebra is spanned by three generators $\{J_\theta\,,J_L=b_L^\dag+b_L\,,J_R=b_R^\dag+b_R \}$, which obey
\be [J_{L(R)}\,,J_\theta]=J_{R(L)}\,,\quad [J_L\,,J_R]=0  \,.\ee
The algebra turns out to be a subalgebra of two commuting $\mathfrak{sl}(2\,,R)$ algebras.

The time-dependent TC state is produced by
\ba\begin{aligned}
|\psi_{TC}(t)\rangle&=e^{-i\bar H t}\, |\mathrm{TC}\rangle\\
&=e^{-i\bar H t} \math{D}(\a,\g;\,\b)e^{i\bar H t}e^{-i\bar H t} |\mathrm{TFD}\rangle\\
&\equiv \math{D}(\a,\g;\,\b,t)|\psi_{TFD}(t) \rangle\,,
\end{aligned}\label{tcstime}\ea
where
\bea
\math{D}(\a,\g;\,\b,t)&\equiv& e^{-i\bar H t} \math{D}(\a,\g;\,\b)e^{i\bar H t}\nn\\
&=&e^{-i\bar H t} D(\hat\a\,,\hat\g)e^{i\bar H t}\nn\\
&=&\mathcal{D}(\hat\alpha\,,\hat\gamma;\,t)\,.
\eea
Thus, a time-dependent TC state is again related to a time-dependent CT state by
\be\label{connection2} |\psi_{TC}(\alpha\,,\gamma;\,t)\rangle=|\psi_{CT}(\hat\alpha\,,\hat\gamma;\,t)\rangle \,,\ee
which generalizes the static counterpart (\ref{connection}). According to this relation, the complexity for a TC state can always be obtained from that of a corresponding CT state
\be\label{cp2} \mathcal{C}_{TC}(\alpha\,,\gamma)=\mathcal{C}_{CT}(\hat{\alpha}\,,\hat{\gamma}) \,.\ee
To avoid redundancy, we will focus on the CT state in the computation of circuit complexity.

\section{Circuit complexity for generalised coherent states}\label{cma}
In this section, we will adopt the covariance matrix approach to calculate the circuit complexity for the CT state. However, the original approach developed for the TFD state \cite{Chapman:2018hou} cannot be applied to our case directly since now our target state is non-Gaussian. One has from (\ref{timetfd}) and (\ref{ctstime})
\ba\begin{aligned}
|\y_T\rangle=|\psi_{CT}(t)\rangle&=e^{-i\lambda_a\xi^a}e^{-\fft{i}{2}k_{ab}^{(0)}\xi^a\xi^b}|0\rangle_L|0\rangle_R\,.
\end{aligned}\label{targetct}\ea
Note that an overall phase factor $e^{-i\omega t/2}$ has been dropped since it does not play any role in our discussions. The non-Gaussianity of the state implies a non-vanishing one-point function of the state. To deal with this and derive the complexity, we will generalize the covariance matrix approach appropriately.

\subsection{Covariance matrix approach: generalisations and complexity}
Considering a quantum system with canonical coordinates $\x^a\equiv(q_1,\cdots,q_N,p_1,\cdots,p_N)$, one has the commutation relations
\ba\label{xaxb}
\left[\x^a,\x^b\right]=i\W^{ab}\,,
\ea
where $\Omega^{ab}$ is an anti-symmetric tensor, given by
\ba
\W^{ab}=\left(
\begin{array}{cc}
  0 & I \\
  -I & 0
\end{array}
\right)\,.
\ea
In the following, we will use $\Omega^{ab}$ to raise indices for tensors, for example
\ba
T_{...c}{}^a{}_{d...}\equiv \W^{ab}T_{...cbd...}\,.
\ea

It was established in \cite{Chapman:2018hou} that for a pure Gaussian state which has a vanishing one-point function, it is completely characterized by its symmetric two-point function, usually referred to as the covariance matrix
\be G^{ab}=\frac{1}{2}\langle \y|\{\x^a, \x^b\}|\y\rangle\,, \label{Gab}\ee
where the $1/2$ factor is introduced for our later purpose. However, for the coherent state and its thermal generalisations, the one-point function is non-vanishing and hence the covariance matrix is not enough to characterize the state.

From (\ref{targetct}), a CT state is connected to the reference state $\bra{\psi_R}$, either Gaussian or non-Gaussian, by two kinds of unitary transformations
\ba
\bra{\y_T}\equiv \hat U \bra{\y_R}= e^{-i\l_a \x^a} e^{-\frac{i}{2}k_{ab}\x^a \x^b}\bra{\y_R}\,.
\ea
The first unitary $ e^{-i\l_a \x^a}$ produces a translation for the canonical variables $\xi^a$ and hence breaks the Gaussianity of the target state. It turns out that to compute the circuit complexity for such a non-Gaussian state, we need to enlarge the covariance matrix (\ref{Gab}) properly. For this purpose, we first introduce the one-point function
\ba\label{onepo}
 \varphi^a \equiv \langle \y|\x^a|\y\rangle\,.
\ea
and then study how the one-point function $\varphi^a$ and the symmetric two-point function $G^{ab}$ transform under the unitary $\hat{U}$.

By simple calculations, we find the following relations
\ba\begin{aligned}
&\left[i\l_b\x^b\,,\x^a\right]=\l^a\,, \hspace{3ex}\left[\fft{i}{2} k_{bc}\x^b\x^c\,,\x^a\right]=K^a{}_b\x^b\,,
\end{aligned}\ea
where
\be \lambda^a\equiv \Omega^{ab}\lambda_b\,,\qquad K^a_{\,\,\,b}\equiv \Omega^{ac}k_{cb} \,.\ee
Notice that $K^{a}_{\,\,\,b}=K_b^{\,\,a}$. Making use of the BCH formula, we get
\bea
&&e^{i\lambda_b \xi^b}\,\xi^a\, e^{-i\lambda_b \xi^b}=\xi^a+\lambda^a\,,\nn\\
&&e^{\fft{i}{2}k_{bc} \xi^b \xi^c}\,\xi^a\, e^{-\fft{i}{2}k_{bc} \xi^b \xi^c}=M^a_{\,\,\,b}\xi^b\,,
\eea
where $M\equiv e^K $. Combining the above results, we are led to
\ba\label{xitrans}
\hat{U}^\dag \x^a \hat{U}=M^a{}_b\x^b+\l^a\,.
\ea
 Interestingly, the above transformation for the canonical variables is similar to the Poincar\'{e} transformation for the Minkowski spacetime. The first unitary $e^{-i\lambda_a \xi^a}$ generates a $\mathbb{R}^{2N}$ translation whilst the second one $e^{-\fft{i}{2}k_{ab}\xi^a \xi^b}$ forms a Lie group $\mathrm{Sp}(2N\,,\mathbb{R})$, playing a role similar to the rotation (boost) in the  Minkowski space. Thus, the full group generated by the unitary $\hat U$ has a structure similar to that of the Poincar\'{e} group, given by the semiproduct of $\mathbb{R}^{2N}$ by the transformations $\mathrm{Sp}(2N\,,\mathbb{R})$
\be\label{group} \mathbb{R}^{2N} \rtimes \mathrm{Sp}(2N\,,\mathbb{R}) \,.\ee
It follows that under the unitary $\hat U$ the one-point function transforms as
\ba
\varphi'^a = \langle \y'|\x^a|\y'\rangle=M^a{}_b\varphi^b+\l^a\,,
\ea
whilst the symmetric two-point function behaves as
\ba\begin{aligned}
G'^{ab}&=\frac{1}{2}\langle \y'|\{\x^a, \x^b\}|\y'\rangle\\
&=\frac{1}{2}\langle \y|\hat{U}^\dag\{\x^a, \x^b\}\hat{U}|\y\rangle\\
&=\frac{1}{2}\langle \y|\{\hat{U}^\dag\x^a\hat{U}, \hat{U}^\dag\x^b\hat{U}\}|\y\rangle\\
&=\frac{1}{2}\langle \y|\{M^a{}_c\x^c+\l^a, M^b{}_d\x^d+\l^b\}|\y\rangle\\
&=M^a{}_c G^{cd} M^b{}_d+M^a{}_c\vp^c\l^b+M^b{}_d\vp^d\l^a+\l^a\l^b\,.
\end{aligned}\ea
In matrix language, the above results can be expressed simply as
\ba\begin{aligned}\label{bG}
\vp'&=\vp M^T+\l\,,\\
G'&=M G M^T+ \l^T \vp M^T+M\vp^T \l+ \l^T \l\,.
\end{aligned}\ea
This is a general transformation that connects two generally non-Gaussian states. For two Gaussian states, it reduces to $\vp'=\vp=0\,,G'=M G M^T$, consistent with the result in \cite{Chapman:2018hou}.

Based on the transformation (\ref{bG}), we introduce an extended covariance matrix
\ba\label{excm}
\bar{G}=\left(
\begin{array}{cc}
  G & \vp^T \\
  \vp & 1
\end{array}
\right)\,,
\ea
so that the transformation (\ref{bG}) can be expressed in a more compact form
\ba\label{bG1}
\bar{G}'=\math{U} \bar{G} \math{U}^T\,,
\ea
where
\ba\label{enlargedU}
\math{U}=\left(
\begin{array}{cc}
  M & \l^T \\
  0 & 1
\end{array}
\right)\,,
\ea
forms a matrix representation for our unitary group.

The quantum circuit that connects the reference state $\bra{\y_R}$ to the target state $\bra{\y_T}$ will be constructed by a series of elements $\math U$. By definition, the complexity is given by the length of the minimal geodesic in the group manifold. However, the geodesic may not be unique and each of them may have a different length, as shown in \cite{Jefferson:2017sdb, Chapman:2018hou}. The reason is that there exists a stabilizer subgroup $V$ for the reference state, i.e., $\forall\ \math{U}_V\in  V$, which leads to
\ba
\math{U}_V\bra{\y_R}=e^{\math{B}_V}\bra{\y_R}=\bra{\y_R}\,,
\ea
where $\math{B}_V$ is the generator of the subalgebra $\mathcal{V}$. This implies that if a unitary $e^{\mathcal{A}}$ (or a geodesic $y(t)=e^{t\mathcal{A}}$ in the group manifold) connects the reference state to the target state, there will be a lot of unitaries (or geodesics) achieving the same goal because of
\ba
e^{\math{A}}e^{\math{B}_V}\bra{\y_R}=e^{\math{A}}\bra{\y_R}=\bra{\y_T}\,.
\ea
Of course, the unitaries $\mathcal{U}_V$ are redundant in the construction of the target state. In other words, if a geodesic that connects the reference state to the target state has unitaries $\mathcal{U}_V$, it will not have the minimal length. This in turn tells us that the optimal geodesic definitely should not have any unitary belonging to the stabilizer subgroup.

To find the optimal geodesic, we first define the inner product on the Lie algebra of the transformation group
\ba
\langle \math{A},\math{B}\rangle =\text{Tr}\lf(\math{A} \bar{G}_R\math{B}^T \bar{g}_R\rt)
\ea
where $\math{A}, \math{B}$ are two infinitesimal generators, $\bar{G}_R$ is the extended covariant metric associated to the reference state and $\bar{g}_R$ is its inverse matrix.
Next we define the horizontal subspace that is transverse to the stabilizer subalgebra
\ba\label{horizonal1}
\math{H}:=\{\math{A}\in \mathcal{G} |\langle \math{A},\math{B}_V\rangle=0,\forall\ \math{B}_V \in \math V\}\,.
\ea
Clearly, a horizontal generator $\mathcal{A}\in  \mathcal{H}$ does not belong to the stabilizer subalgebra and vice versa. However, it does not mean that the Lie algebra $\mathcal{G}$ associated to the transformation group must be split as $\mathcal{G}= \mathcal{H}\oplus \mathcal{V}$. It is still possible that some of the generators do not belong to the product space $\mathcal{H}\oplus \mathcal{V}$, depending on the group structure as well as how we choose the reference state and the target state. It has been shown in \cite{Chapman:2018hou} that when both $\bra{\psi_R}$ and $\bra{\psi_T}$ are Gaussian, the transformation group is $\mathrm{Sp}(2N\,,\mathbb{R})$ and the Lie algebra can indeed split into the product form $\mathfrak{sp}(2N\,,\mathbb{R})=\mathcal{H}\oplus \mathcal{V}$. In this case, the optimal geodesic will be generated by a horizontal generator. However, in our case, the target state is non-Gaussian and the group structure has been enlarged as well as the Lie algebra. It turns out that the above decomposition for our algebra becomes invalid. Nevertheless, as we will show soon,  the extra generator in the algebra is trivial as long as the reference state is Gaussian. It does not connect the reference state to the target state so that the optimal geodesic in our case will still be generated by a horizontal generator, i.e., $y(t)=e^{t\math{A}}$, where $\math{A}\in \math H$. The complexity of the target state will be given by
\ba\label{cplex}
\math{C}\big(\bra{\y_R}\to\bra{\y_T}\big)= \|\math{A}\|,
\ea
where we have taken the cost function to be $F_2$.
Therefore, evaluating the complexity is equivalent to finding the horizonal generator $\math{A}$ such that
\ba
\bra{\y_T}=e^{\math{A}}\bra{\y_R}\,.
\ea
The uniqueness of the generator will be guaranteed by our derivations below.

In the remaining of this paper, to derive the complexity, we choose the reference state to be a Gaussian state, which has the extended covariance matrix
\ba
\bar{G}_R=\left(
\begin{array}{cc}
  G_R & 0 \\
  0 & 1
\end{array}
\right)\,.
\ea
In this case, the stabilizer subgroup can be defined by
\ba
V:=\{\math{U}_V\in G| \math{U}_V \bar{G}_R\math{U}_V^T=\bar{G}_R\}\,.
\ea
Consequently, its Lie algebra satisfies
\ba\label{OBV}
\math{V}=\{\math{B}_V\in \math{G}| \lf(\math{B}_V\bar{G}_R\rt)^T=-\math{B}_V\bar{G}_R\}\,.
\ea
According to (\ref{horizonal1}), this leads to
\be \text{Tr}\big(\math A \math B_V\big)=0 \,.\ee
It implies that the horizontal generators obey
\ba\label{OH0}
\math{H}=\{\math{A}\in \math{G}| (\math A \bar{G}_R)^T=\math A \bar{G}_R\}\,.
\ea
From Eq.(\ref{OBV}), the Lie algebra of the stabilizer subgroup can be expressed as
\ba\label{statrue}
\math{V}=\left\{\math{B}\in \math{G}\left| \math{B}=\left(
\begin{array}{cc}
  O_\math{B} & 0 \\
  0 & 0
\end{array}
\right)\right. \quad \text{with}\quad \lf(O_\math{B}G_R\rt)^T=-O_\math{B}G_R\right\}\,,
\ea
whilst Eq.(\ref{OH0}) implies that the horizontal generators satisfy
\ba\label{OH00}
\math{H}=\left\{\math{A}\in \math{G}\left| \math{A}=\left(
\begin{array}{cc}
  O_\math{A} & \n_A^T \\
  0 & 0
\end{array}
\right)\right. \quad \text{with}\quad \lf(O_\math{A}G_R\rt)^T=O_\math{A}G_R\right\}\,.
\ea
Notice that though $O_\math{A}\oplus O_\math{B}$ gives rise to the full subalgebra $\mathfrak{sp}(2N\,,\mathbb{R})$, the decomposition $\mathcal{G}= \mathcal{H}\oplus \mathcal{V}$ does not hold because an extra generator remains. One finds
\be\label{productalgebra}
\mathcal{G}= \mathcal{H}\oplus \mathcal{V}\oplus
\begin{pmatrix}
0&0\\
0&1
\end{pmatrix}\,.
\ee
However, according to the transformation (\ref{bG1}), the last generator just transforms a unity to a unity and hence is trivial. Therefore, in our case the optimal geodesic will still be generated by a horizontal generator.

For later convenience, we parameterize the horizontal generator that connects the reference state and the target state to be
\ba\label{AO}
\math{A}=\left(
\begin{array}{cc}
  K & \n^T \\
  0 & 0
\end{array}
\right)\,,
\ea
where $\nu$ is a vector which will be determined later and $K\in \mathfrak{sp}(2N\,,\mathbb{R})$ obeying
\ba\label{OH}
\big( K G_R \big)^T=K G_R\quad \Rightarrow \quad K^T=g_R K G_R\,.
\ea
Evaluating the exponential of the horizonal generator yields
\ba\label{AO2}
\math{U}=e^{\math{A}}=\exp\lf[\left(
\begin{array}{cc}
  K & \n^T \\
  0 & 0
\end{array}
\right)\rt]=\left(
\begin{array}{cc}
  M & \chi^T \\
  0 & 1
\end{array}
\right)\,.
\ea
Comparing to (\ref{enlargedU}), one immediately finds $\chi=\lambda$. The vector $\nu$ is determined by
\be \nu^T=\big(M-I\big)^{-1}K \l^T \,.\label{vectornu}\ee
It is worth emphasizing that a nontrivial vector $\nu$ is essentially determined by a nonvanishing one-point function for the target state. On the contrary, when the target state is Gaussian, one has $\vp=0=\l$, leading to $\nu=0$.

Now it is straightforward to derive the complexity
\ba\label{cplex2}\begin{aligned}
\math{C}^2(\bra{\y}_R\to\bra{\y}_T)&=\|\math{A}\|^2\\
&=\text{Tr}\lf(\math{A}^T\bar{g}_R\math{A}\bar{G}_R\rt)\\
&=\text{Tr}\lf[\left(
\begin{array}{cc}
  K^Tg_R K G_R & K^Tg_R\n^T \\
  \n g_RKG_R & \n g_R\n^T
\end{array}
\right)\rt]\\
&=\text{Tr}\big (K^2 \big)+\n g_R\n^T\,.
\end{aligned}\ea
However, we have not solved the generator $K$ in terms of the information for $|\y_R\rangle$ and $|\y_T\rangle$. According to the transformation (\ref{bG}), the one-point function and the covariance matrix of the target state are given by (recall that the reference state is Gaussian)
\bea\label{gOG}
&&\varphi=\chi=\lambda\,,\nn\\
&&G_T=M G_R M^T+ \l^T \l\,.
\eea
On the other hand, using (\ref{OH}), one has
\ba
M^T=e^{K^T}=e^{g_R K G_R}=g_R M G_R\,.
\ea
Then from \eq{gOG}, one finds
\ba\label{EOD}
M^2=e^{2K}=\D_T^{(0)}\,,\quad  \D_T^{(0)}\equiv \D_T- \l^T \l\, g_R\,,
\ea
where $\D_T\equiv G_T g_R$ is called the relative covariance matrix \cite{Chapman:2018hou}. This solves the generator $K$ and its matrix exponential. However, it is worth emphasizing that for any given Gaussian reference state, the quantity $\D_T^{(0)}$ as well as the generator $K$ does not really depend on the non-Gaussianity of the target states (this is not hard to understand since $K$ generates rotations only for the canonical variables $\xi^a$, instead of translation). In sec.\ref{gaussion}, we will explicitly show that $\D_T^{(0)}$ is nothing else but simply the relative covariance matrix for the ground thermofield double state, namely
\be \D_T^{(0)}=\D_T(\lambda=0) \,.\label{rrcovmatrix}\ee

The complexity turns out to be
\bea \math{C}^2&=&\mathcal{C}^2_{(0)}+\n g_R\n^T\nn\\
&=&\frac{1}{4}\text{Tr}\Big[\Big(\log \D_T^{(0)} \Big)^2 \Big]+\n g_R\n^T \,.\label{cpgene}\eea
In general, the result gives rise to the complexity between a non-Gaussian target state and a Gaussian reference state.
However, it also establishes the relation between the complexity for non-Gaussian target states and that of the ground state. In fact, for any Gaussian reference state, one has according to (\ref{generalCM})
 \be \math{C}^2-\mathcal{C}^2_{(0)}=\n g_R\n^T\geq 0\,,\label{compareformula}\ee
 where the equality is taken when the target state becomes purely Gaussian. The result implies that for a same Gaussian reference state, the complexity for an excited state is always larger than that of the ground state.

Last but not least, despite that our target state has been specified in (\ref{targetct}), we can still choose different Gaussian reference states. With different choices, the difficulty for the calculations and the results will be significantly different, as will be shown below.

\subsection{Dirac vacuum as the reference state}\label{dvatrs}

In this subsection, we will calculate the complexity analytically by choosing the Dirac vacuum as the reference state, namely $\bra{\y_R}=\bra{0}_L \bra{0}_R$.

By definition, the covariance matrix for this reference state can be evaluated as
\ba\begin{aligned}
G_R=\left(
\begin{array}{cccc}
 \frac{1}{2m\w} & 0 & 0 & 0 \\
 0 & \frac{1}{2m\w} & 0 & 0 \\
 0 & 0 & \frac{m\w}{2} & 0 \\
 0 & 0 & 0 & \frac{m\w}{2}
 \end{array}\right)\,.
\end{aligned}\ea
It turns out that for this particular reference state, the unitary connecting $\bra{\y_R}$ to $\bra{\y_T}$ has already been given in (\ref{targetct}) and hence the generator $K$ can be read off directly. One has $k_{ab}=k_{ab}^{(0)}$ and hence $K^{a}_{\,\,\,b}=\Omega^{ac}k_{cb}^{(0)}$. The matrix form is given by
\ba
 K=
\theta\left(
\begin{array}{cccccccc}
 0 & \cos \w t & 0  & -\frac{\sin\w t}{m\w} \\
 \cos \w t & 0 & -\frac{\sin\w t}{m\w} & 0  \\
 0 & -m\w\sin\w t & 0 & -\cos \w t \\
 -m\w\sin\w t & 0 & -\cos \w t & 0 \\
\end{array}
\right)\,.
\ea
Next we derive its matrix exponential $M=e^K$. We find that the generator $K$ can be diagonalized by a matrix $S$, i.e., $K=S K_0 S^{-1}$ with $K_0=\text{diag}\{-\theta,-\theta,\theta,\theta\}$ and
\ba
S=
\left(
\begin{array}{cccc}
 \frac{\csc \omega t}{m\w} & -\frac{\cot \omega t}{m\w} & -\frac{\csc\omega t}{m\w} & -\frac{\cot \omega t}{m\w} \\
 -\frac{\cot \omega t}{m\w} & \frac{\csc \omega t}{m\w} & -\frac{\cot \omega t}{m\w} & -\frac{\csc\omega t}{m\w} \\
 0 & 1 & 0 & 1 \\
 1 & 0 & 1 & 0 \\
\end{array}
\right)\,.
\ea
It is easy to see that $\text{Tr}\big( K^2\big)=\text{Tr}\big( K_0^2\big)=4\theta^2$. The transformation matrix $M$ can be evaluated as
\ba
M=e^{ K}=e^{S K_0 S^{-1}}=S e^{K_0} S^{-1}\,,
\ea
where $e^{ K_0}$ is a diagonalized matrix $e^{ K_0}=\text{diag}\{e^{ -\q},e^{- \q},e^{ \q},e^{ \q}\}$. We deduce
\ba\label{M1}
M=
\left(
\begin{array}{cccc}
 \cosh \theta  & \cos  \omega t \sinh \theta  & 0 & -\frac{\sin  \omega t \sinh \theta}{m\w}  \\
 \cos  \omega t \sinh \theta  & \cosh \theta  & -\frac{\sin  \omega t \sinh \theta}{m\w}  & 0 \\
 0 & -m\w\sin \omega t \sinh \theta  & \cosh \theta  & -\cos  \omega t \sinh \theta  \\
 -m\w\sin \omega t \sinh \theta  & 0 & -\cos  \omega t \sinh \theta  & \cosh \theta  \\
\end{array}
\right)\,.
\ea
On the other hand, we have
\ba
\l^a=\Omega^{ab}\lambda_b=\big(\l_{p_L}\,,\l_{p_R}\,,-\l_{q_L}\,,-\l_{q_R}\big)\,,
\ea
where $\lambda_a$'s have been specified in (\ref{CTSlambda}). With these results in hand, it is straightforward to evaluate the vector $\nu$ using the formula (\ref{vectornu}). We obtain
\ba
\nu^T=\fft{\theta}{2}\left(
\begin{array}{c}
   \l_{p_L}\coth{\big(\fft{\theta}{2}\big)}-\l_{p_R}\cos{(\w t)}-\l_{q_R}\fft{\sin{(\w t)}}{m\w}  \\
   \l_{p_R}\coth{\big(\fft{\theta}{2}\big)}-\l_{p_L}\cos{(\w t)}-\l_{q_L}\fft{\sin{(\w t)}}{m\w}\\
   -\l_{q_L}\coth{\big(\fft{\theta}{2}\big)}-\l_{q_R}\cos{(\w t)}+\l_{p_R} m\w\sin{(\w t)}\\
   -\l_{q_R}\coth{\big(\fft{\theta}{2}\big)}-\l_{q_L}\cos{(\w t)}+\l_{p_L}m\w\sin{(\w t)}
\end{array}
\right)\,.
\ea
Finally, using (\ref{cplex2}) we read the complexity
\be\label{cpcts} \mathcal{C}=\theta\, \text{csch}\big(\ft{\theta}{2} \big) \sqrt{\big(|\alpha|^2+|\gamma|^2+2 \big)\cosh\theta+2(\Im\alpha \Im\gamma-\Re\alpha \Re\gamma)\sinh\theta-2} \,.\ee
Remarkably, the result is independent of time! Notice that the complexity only depends on the quadratic polynomials of the real and imaginary parts of $\alpha$ and $\gamma$. Thus, we have
\be\label{cprelation} \mathcal{C}(\alpha\,,\gamma)=\mathcal{C}(\gamma\,,\alpha)=\mathcal{C}(-\alpha\,,-\gamma)=\mathcal{C}(\alpha^*\,,\gamma^*) \,.\ee
In fact, for some special cases, the complexity enjoys even more symmetries.

\textbf{a) Single excitation ($\gamma=0$):} At first, we would like to consider single excitations. Without loss of generality, we set $\gamma=0$. Then, we have
\be
\mathcal{C}_a(\alpha\,,0)=\theta\, \text{csch}\big(\ft{\theta}{2} \big) \sqrt{\big(|\alpha|^2+2 \big)\cosh\theta-2} \,.
\ee
It is immediately seen that the complexity depends only on $|\alpha|$, implying that for any $\alpha'=|\alpha|e^{i\varphi}$, $\mathcal{C}_a(\alpha'\,,0)=\mathcal{C}_a(\alpha\,,0)$. This is much stronger than the relation (\ref{cprelation}).

\textbf{b) Two equal excitations ($\gamma=\alpha$):}
Secondly, we move to the two equal excitation case. The Eq. (\ref{cpcts}) gives rise to
\be
 \mathcal{C}_b(\alpha\,,\alpha)=\theta\, \text{csch}\big(\ft{\theta}{2} \big) \sqrt{2\big(|\alpha|^2+1 \big)\cosh\theta-\left(\alpha^2+\alpha^{\ast2} \right)\sinh\theta-2} \,.
 \ee
 One easily finds $\mathcal{C}_b(\alpha\,,\alpha)=\mathcal{C}_b(-\alpha\,,-\alpha)=\mathcal{C}_b(\alpha^\ast\,,\alpha^*)$. It is interesting to compare $\c_{b}$ with $\c_{a}$ and $2\c_{a}$ for a same eigenvalue $\alpha$. We have
 \bea
 \c_{b}^2-\c_{a}^2&=&\theta^2\mathrm{csch}(\ft{\theta}{2})^2\Big(|\alpha|^2\cosh\theta+2(\Im\alpha^2-\Re\alpha^2)\sinh\theta\Big)\,,\\
(2\c_{a})^2-\c_{b}^2&=&2\theta^2\mathrm{csch}(\ft{\theta}{2})^2\Big((3+|\alpha|^2)\cosh\theta+(\Re\alpha^2-\Im\alpha^2)\sinh\theta-3\Big)\,.
 \eea
 Recall $\tanh \theta=e^{-\beta \omega/2}>0$ which implies $\theta>0$. However, since $\Re\alpha$ and $\Im\alpha$ are arbitrary, the sign of the above two equations are not fixed. It implies that the two modes are highly entangled and have a strong temperature dependence.

\textbf{c) Two opposite excitations ($\gamma=-\alpha$):} Next, we consider the target state with two opposite excitations. Using Eq. (\ref{cpcts}), we obtain
\be
 \mathcal{C}_c(\alpha\,,-\alpha)=\theta\, \text{csch}\big(\ft{\theta}{2} \big) \sqrt{2\big(|\alpha|^2+1 \big)\cosh\theta+\left(\alpha^2+\alpha^{\ast2} \right)\sinh\theta-2} \,.
 \ee
Similar to the two equal excitation case, we find $\mathcal{C}_c(\alpha\,,-\alpha)=\mathcal{C}_c(-\alpha\,,\alpha)=\mathcal{C}_c(\alpha^\ast\,,-\alpha^*)$. Furthermore, we have
\be \mathcal{C}_c(\alpha\,,-\alpha)^2-\mathcal{C}_b(\alpha\,,\alpha)^2=8\theta^2\coth{(\ft{\theta}{2})}\big(\Re\alpha^2-\Im\alpha^2 \big) \,,\ee
which is positive for $|\Re\alpha|>|\Im\alpha|$ and negative for $|\Re\alpha|<|\Im\alpha|$, independent of the temperature.

\textbf{d) Two conjugated excitations ($\gamma=\alpha^\ast$):} At last, let\rq{}s turn to the two conjugated excitation case. The complexity is given by
\be
 \mathcal{C}_d(\alpha\,,\alpha^*)=\theta\, \text{csch}\big(\ft{\theta}{2} \big) \sqrt{2\big(|\alpha|^2+1 \big)\cosh\theta-2|\alpha|^2 \sinh\theta-2} \,.
 \ee
  It is interesting to note that the result depends only on $|\alpha|$, similar to the single excitation case.  Moreover, comparing $\c_d$ with $\c_c$ and $\c_b$ for a same $\alpha$, we find
 \bea
 \c_d(\alpha\,,\alpha^*)^2-\c_c(\alpha\,,-\alpha)^2&=&-8\Re\alpha^2 \theta^2\coth{(\ft{\theta}{2})}\,,\\
  \c_d(\alpha\,,\alpha^*)^2-\c_b(\alpha\,,\alpha)^2&=&-8\Im\alpha^2\theta^2\coth{(\ft{\theta}{2})}\,,
 \eea
 which are both negative for any non-vanishing $\alpha$ as well as the temperature.

 However, it is worth emphasizing that the above relations between the complexity for different types of excitations strongly depend on the reference state. We will turn to this point again in sec.\ref{twomode} (see the last paragraph for two conjugated excitations).


\subsection{More general reference state}\label{gaussion}

While we have chosen the reference state to be a Gaussian state, it is not necessarily to be the Dirac vacuum. A more general Gaussian state could be characterized by a covariance matrix
\ba\begin{aligned}
G_R=\left(
\begin{array}{cccc}
 \frac{1}{2\eta m\w} & 0 & 0 & 0 \\
 0 & \frac{1}{2\eta m\w} & 0 & 0 \\
 0 & 0 & \frac{\eta m\w}{2} & 0 \\
 0 & 0 & 0 & \frac{\eta m\w}{2}
 \end{array}\right)\,,
\end{aligned}\label{generalCM}\ea
where $\eta>0$ and the state is no longer the Dirac vacua when $\eta\neq 1$. As a consequence, the generator $K$ cannot be read off from (\ref{targetct}) any longer. Instead, we shall solve it in terms of the relative covariance matrix using the formula (\ref{EOD}). For this purpose, we need to derive the covariance matrix $G_T$ for the target state at first. According to the definition (\ref{Gab}), $G_T$ should be independent of the reference state. Thus, we can calculate it from the transformation \eq{bG} by choosing a particular reference state, namely the Dirac vacua. We obtain
\ba\label{comatrix}
G_T=\left(
\begin{smallmatrix}
  \lambda_{p_L}^2+\fft{\cosh2\q}{2m\w} & \,\,\,\,\,\l_{p_L}\l_{p_R}+\fft{\cos \w t  \sinh 2\q}{2m\w} & -\l_{p_L}\l_{q_L} & -\l_{p_L}\l_{q_R}-\ft 12\sin \w t \sinh 2\q \\
   \cdots & \l_{p_R}^2+\fft{\cosh2\q}{2m\w} &  -\l_{p_R}\l_{q_L}-\ft12 \sin \w t \sinh2\q & -\l_{p_R}\l_{q_R} \\
  \cdots & \cdots & \l_{q_L}^2+\ft 12 m\w \cosh2\q  & \l_{q_L}\l_{q_R}-\ft12 m\w \cos\w t\sinh2\q  \\
    \cdots & \cdots & \cdots  & \l_{q_R}^2+\ft12 m\w \cosh2\q
\end{smallmatrix}
\right)\,.
\ea
Furthermore, substituting the result into (\ref{EOD}) yields
\ba\label{rrcomatrix}
\D_T^{(0)}=\left(
\begin{smallmatrix}
 \eta\cosh2\q & \eta\cos \w t  \sinh 2\q & 0 & -\ft{\sin \w t \sinh 2\q}{\eta m\omega} \\
   \eta\cos \w t  \sinh 2\q & \eta\cosh2\q &  -\ft{\sin \w t \sinh2\q}{\eta m \omega} & 0 \\
  0 & -\eta m\w \sin\w t\sinh 2\q &  \ft{\cosh2\q}{\eta}  & -\ft{\cos\w t\sinh2\q}{\eta}  \\
    -\eta m\w \sin\w t\sinh 2\q & 0 & -\ft{\cos\w t\sinh2\q}{\eta}  &  \ft{\cosh2\q}{\eta}
\end{smallmatrix}
\right)\,.
\ea
It is immediately seen that $\D_T^{(0)}$ does not rely on the translation generator $\lambda$. It simply gives the relative covariance matrix for the ground thermodfield double state.

However, now it is of great difficult to solve the generator $K$ and the vector $\nu$ analytically. Nevertheless, since we have extracted the covariance matrix $G_T$ for the target state, it is straightforward to solve them numerically. This is a purely algebraic problem: the generator $K$ can be solved from (\ref{EOD}) and then the vector $\nu$ will be determined by (\ref{vectornu}). Finally, it is straightforward to obtain the complexity through (\ref{cplex2}) or (\ref{cpgene}).

According to these considerations as well as the expression (\ref{CTSlambda}) for the translation generator $\lambda$, we can reasonably speculate that in general the complexity for the CT states is a time periodic function and may take the form of
\be \mathcal{C}^2=\sum_{n=0}^{N} a_n\sin{(n\omega t+\varphi_n)} \,,\quad \varphi_0=\pi/2\,,\ee
where $N=0\,,1\,,2\,,\cdots$ is some nonnegative integer. The specific value of $N$ as well as the amplitudes $a_n$ and the initial phases $\varphi_n$ for above each term are all functions of the parameters $(\alpha\,,\gamma\,,\beta\,,\eta)$. Indeed, by carefully scanning the parameters space, we find that all of our numerical results can be perfectly fitted by the above formula. This in turn helps us to understand the numerical results better. For example, we can use the fitting formula to precisely test some of our relations for complexity that are read off from numerical results.

 In the following, we will show how to extract some important information about the evolution of complexity via numerical analysis. However, before moving to this, we shall explain some of our notations at first. Since we are particularly interested in the dependence on the eigenvalues, the complexity for the CT states will be denoted by $\mathcal{C}_{(\alpha\,;\gamma)}(t)$. It was understood that the states have eigenvalues $\alpha (\gamma)$ on the left (right) hand side.  However, sometimes it is more convenient for us to label the complexity by using the real and imaginary parts of the eigenvalues directly. We denote $\mathcal{C}_{(\alpha\,;\gamma)}(t)=\mathcal{C}_{(P,Q\,;U,V)}(t)$, where $P\,,Q\,,U\,,V$ take real values and are related to the eigenvalues $\alpha$ and $\gamma$ by
\bea\label{PQMN}
\alpha=P+iQ,\quad\quad\gamma=U+iV.
\eea
For example, $\mathcal{C}_{(1,2;0,0)}(t)$ denotes the time-dependent complexity of the target state with $\alpha=1+2i\,,\gamma=0$, and $\mathcal{C}_{(+,-;0,0)}(t)$ denotes $\mathcal{C}_{(P,Q\,;U,V)}(t)$ with $P>0, Q<0, U=V=0$. We will frequently switch between these two notations. The readers should not be confused.

In addition, we choose the initial time of the evolution to be $t=0$. In our numerical results, we will simply show the evolution for the time regime $t\geq 0$. However, it should be emphasized that there is no difficulty to analytically continue our results to the $t<0$ regime. In fact, some of our formulas that are obtained from the numerical results should be understood in this way, for example $\mathcal{C}_{(\alpha^*\,;\alpha^*)}(t)=\mathcal{C}_{(\alpha\,;\alpha)}(2\pi/\omega-t)$. To avoid redundancy, we will not emphasize this again in the following.

\begin{figure}[ht]
\centering
\includegraphics[width=450pt]{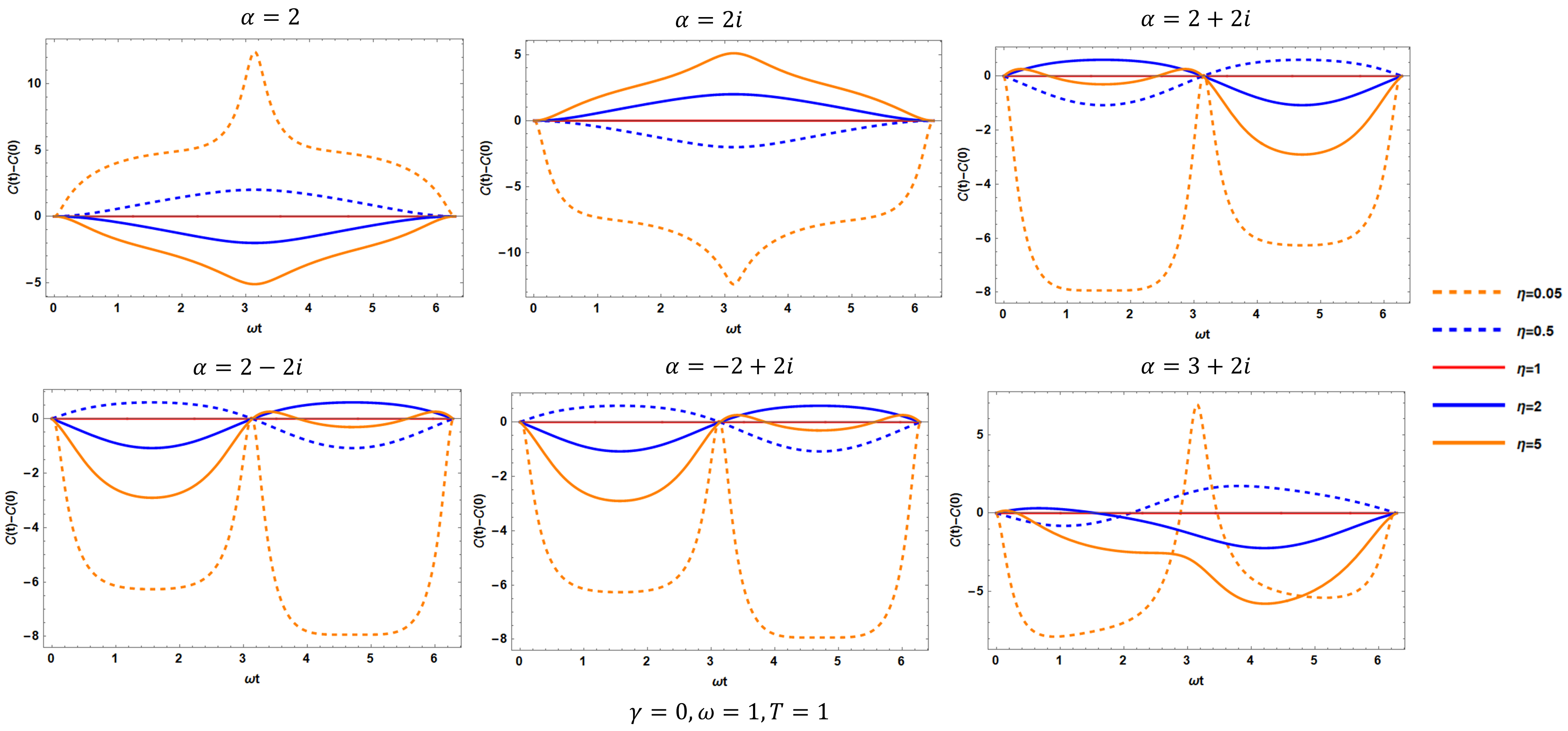}
\caption{{\it The growth of the complexity for single excitations with various $\eta$ and $\alpha$.  }}
\label{single1}
\end{figure}

\subsubsection{Single excitation}
Let us start with a simple case in which the target state is only excited by a single mode. Without loss of generality, we assume $\gamma=0$. Here, we find that it is more convenient for us to study a new quantity
\be
C_{(\alpha\,;\gamma)}(t)=\c_{(\alpha\,;\gamma)}(t)-\c_{(\alpha\,;\gamma)}(0)\,,
\ee
which describes the growth of complexity (GC) from the value at the initial time. It should not be confused with the complexity itself.

In Fig. \ref{single1}, we show the evolution of the GC for single-mode excitations (in the first period) with various $\eta$ and $\alpha$. From the figure, we can read off a lot of interesting features.
\begin{enumerate}
\item First, the (minimal) time period of the GC (or the complexity) is given by $2\pi/\omega$ rather than $\pi/\omega$. The latter is known to be the time period for the ground thermal field double state \cite{Chapman:2018hou}. The difference can be attributed to the non-Gaussianity of our target state, which now has a non-vanishing vector $\nu$. By carefully examining Eq. (\ref{cplex2}), we find that the first term $\mathrm{Tr}(K^2)$ gives lowest order terms as $\cos^2(\omega t)$ or $\sin^2(\omega t)$ while the second term $\n g_R\n^T$ results in terms like $\cos(\omega t)$ or $\sin(\omega t)$. Thus, in our framework, one can generally distinguish a non-Gaussian target state from a Gaussian state using the minimal evolving time period of the complexity.

\item Secondly, the GC is sensitive to the parameters. For example, when only the real part of $\alpha$ mode is excited, $C_{(+,0\,;0,0)}(t)\ge0$ provided $\eta<1$, otherwise $C_{(+,0\,;0,0)}\le0$. On the contrary, if only the imaginary part of $\alpha$ mode is excited, the behaviour will be the opposite.

\item Thirdly, when the eigenvalue $\alpha$ is real or pure imaginary,  the GC (or complexity) is a symmetric function about the axis $t=\pi/\omega$ in the first period. This implies $C_{(\alpha\,;0)}(t)=C_{(\alpha\,;0)}(2\pi/\omega-t)$.

\item For general $\alpha$, there exists an inversion relation $C_{(\alpha\,;0)}(t)=C_{(-\alpha\,;0)}(t)$. The result is valid to the complexity as well because we find initially the complexity obeys $\c_{(\alpha\,;0)}(0)=\c_{(-\alpha\,;0)}(0)=\c_{(\pm\alpha^*\,;0)}(0)$.

\item Interestingly, the GC for two single excitation states with conjugated eigenvalues are symmetric about the axis $t=\pi/\omega$ and hence can be connected by $C_{(\alpha^*\,;0)}(t)=C_{(\alpha\,;0)}(2\pi/\omega-t)$. The relation is valid to the complexity as well because of the above relation for the initial complexity.

\item Finally, from the lower three panels in Fig. \ref{single1}, we conclude that there exists an identity $C_{(\alpha\,;0)}(\pi/\omega)=0$ or $\mathcal{C}_{(\alpha\,;0)}(\pi/\omega)=\mathcal{C}_{(\alpha\,;0)}(0)$, which holds if and only if the eigenvalue takes the form of $\alpha=P(1\pm i)$, where $P$ is a real number. In fact, this can be explained by our expression (\ref{CTSlambda}) for $\lambda$. Because of $\lambda^a=(\sqrt{\ft{2}{m\omega}}\,\Re\alpha\,,0\,,\sqrt{2m\omega}\,\Im\alpha\,,0)$ when $t=0$ and $\lambda^a=(\sqrt{\ft{2}{m\omega}}\,\Im\alpha\,,0\,,-\sqrt{2m\omega}\,\Re\alpha\,,0)$ when $t=\pi/\omega$, one finds that for $\alpha=P(1\pm i)$, $\mathcal{C}_{(P,\pm P\,;0,0)}(\pi/\omega)=\mathcal{C}_{(\pm P,-P\,;0,0)}(0)=\mathcal{C}_{(P,\pm P\,;0,0)}(0)$, where the second equality follows from the above relation for the initial complexity.

\end{enumerate}
\begin{figure}[ht]
\centering
\includegraphics[width=450pt]{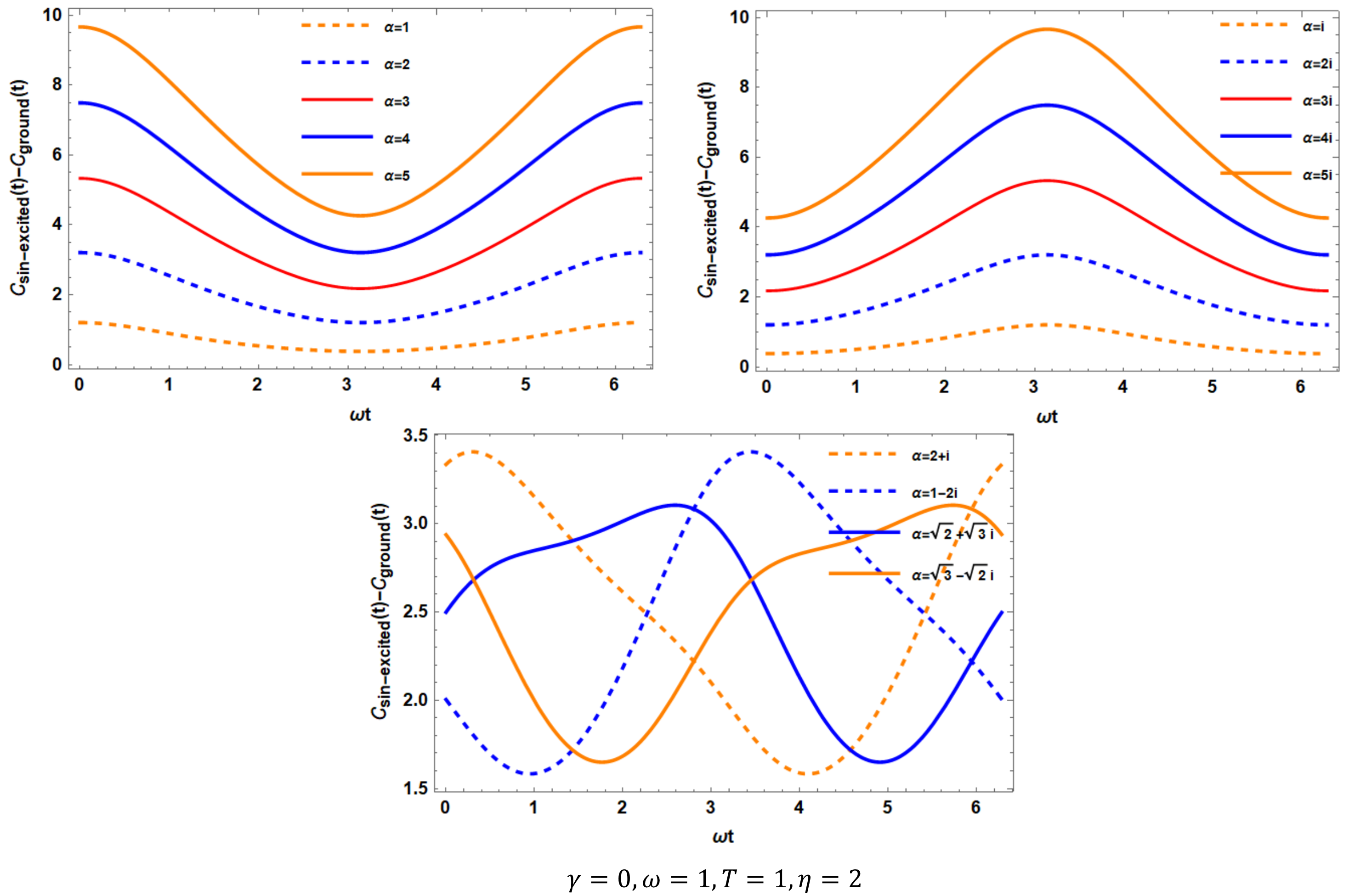}
\caption{{\it The difference of the complexity between a single excitation state and the ground state. In the upper panels, the eigenvalue $\alpha$ is real (left) and pure imaginary (right). In the lower panel, we compare the results for two single excitation states with eigenvalues $\alpha$ and $-i\alpha$.  }}
\label{compare}
\end{figure}
 One may notice that in Fig. \ref{single1}, the red line describing the GC for the ground thermofield double state is sometimes above the other tinctorial lines. However, this does not mean that the complexity of the ground state is larger than that of an excited state, since our pictures simply show the growth of the complexity, rather than the complexity itself. As a matter of fact, the complexity for an excited state at the initial time has already been large enough, which guarantees that it costs more to prepare the state from the reference state, compared to the ground state. To show this, we plot the difference between the complexity of an excited state and the ground state directly in Fig. \ref{compare}. In the upper panels, the eigenvalue $\alpha$ is real in the left panel and is pure imaginary in the right panel while in the lower panel, we concentrate on the results for two single excitation states with eigenvalues $\alpha$ and $-i\alpha$. It is clear that for all these cases, the difference is always positive definite, consistent with our previous argument (\ref{compareformula}). In surprise, from the figure, we also find a translation formula for the complexity itself:
 \begin{enumerate}
 \item For general $\alpha$, the complexity for two single excited states with eigenvalues $\alpha$ and $\pm i\alpha$ obeys $\c_{(\pm i\alpha\,;0)}(t)=\c_{(\alpha\,;0)}(t+\pi/\omega)$ (recall that the minimal time period for the ground state is $\pi/\omega$). It also implies that $\c_{(\pm i\alpha^*\,;0)}(t)=\c_{(\mp i\alpha\,;0)}(2\pi/\omega-t)=\c_{(\alpha\,;0)}(3\pi/\omega-t)$.

 \end{enumerate}
\begin{figure}[ht]
\centering
\includegraphics[width=450pt]{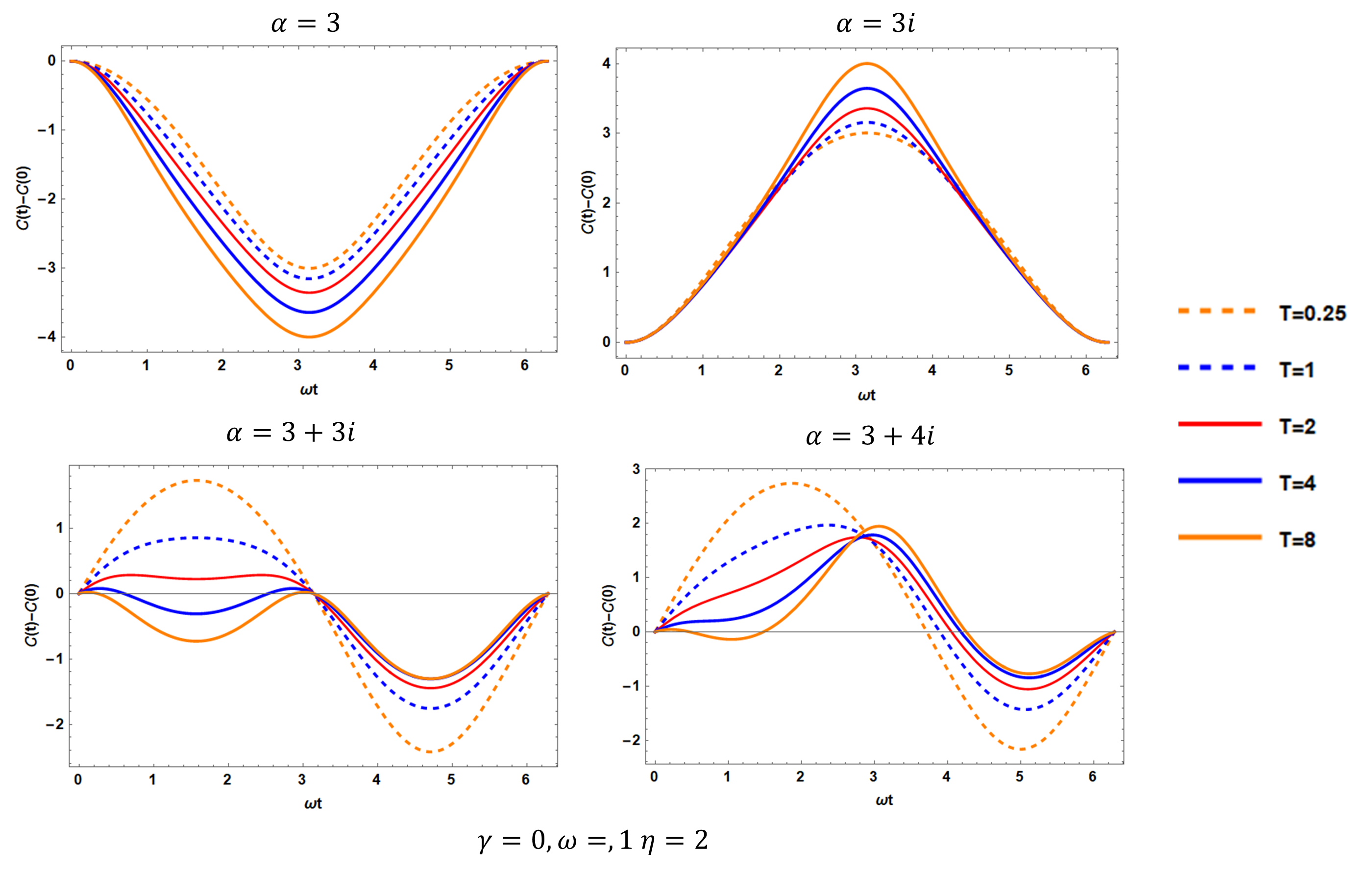}
\caption{{\it The GC for single excitations with various temperatures. }}
\label{single3}
\end{figure}

To proceed, we would like to further study the effects of the temperature on the GC. For this purpose, we fix the other parameters and let $T$ vary and then compare the complexity for the target states at different temperatures. Some of our numerical results are shown in Fig. \ref{single3}. From the figure, we conclude as follows.

\begin{enumerate}
\item From the upper left panel, we observe that $C_{(+,0\,;0,0)}(\pi/\omega)$ takes the minimal value in the first period and $C_{(+,0\,;0,0)}(\pi/\omega)$ decreases as the temperature increases. On the contrary, the upper right panel shows that $C_{(0,+\,;0,0)}(\pi/\omega)$ takes the maximal value in the period and $C_{(0,+\,;0,0)}(\pi/\omega)$ increases as the temperature increases.

\item We rediscover that the GC vanishes at $t=\pi/\omega$ for the eigenvalues $\alpha=P(1\pm i)$ and this point divides the image into two parts. For example, when $\alpha=P(1+i)\,,P>0$, the extreme value decreases as the temperature increases in the first half period $(0, \pi/\omega)$ while in the other half period it behaves in the opposite way, as shown in the lower left panel.

\item As we have mentioned above, $C_{(\alpha\,;0)}(\pi/\omega)\neq0$ when $\alpha\neq P(1\pm i)$. In this case, no obvious laws can be found easily, see the lower right panel.

\end{enumerate}

\subsubsection{Two excitations}\label{twomode}

In this subsubsection, we focus on the two excitation states, which have $\alpha\neq0$ and $\gamma\neq0$.
In the following, we will study three special cases: two equal excitations $\alpha=\gamma$, two opposite excitations $\alpha=-\gamma$ and two conjugated excitations $\alpha=\gamma^\ast$ respectively, to illustrate some main features about the evolution of complexity for two excitation states.

\begin{figure}[ht]
\centering
\includegraphics[width=450pt]{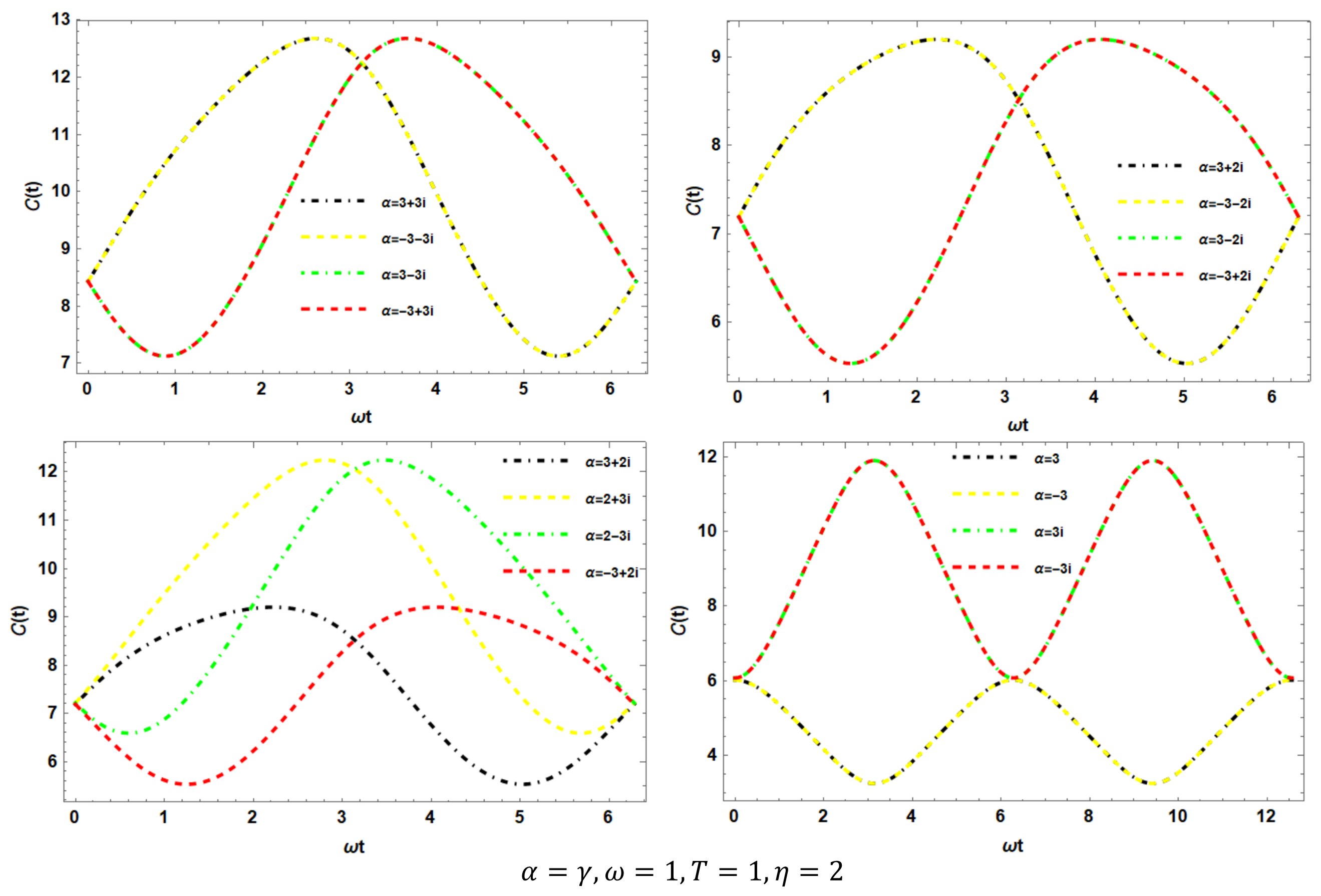}
\caption{{\it The evolution of complexity for two equal excitations with various $\alpha$.}}
\label{twoeq1}
\end{figure}

\textbf{a) Two equal excitations ($\alpha=\gamma$)}

Let us start with two equal excitations. In Fig. \ref{twoeq1}, we show the evolution of complexity for various eigenvalues $\alpha$. From the figure, we find some interesting features.
\begin{enumerate}

\item The first is that for general $\alpha$, there exists a relation $\c_{(\alpha\,;\alpha)}(t)=\c_{(-\alpha\,;-\alpha)}(t)$.

\item From the upper panels, we find that similar to the single excitation case, the complexity for two excited states with conjugated eigenvalues are symmetric about the axis $t=\pi/\omega$ in the first period and hence $\c_{(\alpha^*\,;\alpha^*)}(t)=\c_{(\alpha\,;\alpha)}(2\pi/\omega-t)$.

\item Secondly, from the lower left panel, we observe that the states with a larger absolute value of the imaginary part $|\Im\alpha|$ have a higher peak value of the complexity. It may suggest that the imaginary part of $\alpha$ or $\gamma$ gives greater influence in preparing two equal excited states.

\item In particular, from the lower right panel, we find a new interesting relation $\mathcal{C}_{(i\alpha\,;i\alpha)}^{min}=\mathcal{C}_{(\alpha\,;\alpha)}^{max}$, where $\alpha$ is real. The result implies that $\mathcal{C}_{(i\alpha\,;i\alpha)}(t)\geq \mathcal{C}_{(\alpha\,;\alpha)}(t')$ for any time $t\,,t'$. This strongly supports the above argument.

\end{enumerate}

\textbf{b) Two opposite excitations ($\alpha=-\gamma$). }

\begin{figure}[ht]
\centering
\includegraphics[width=450pt]{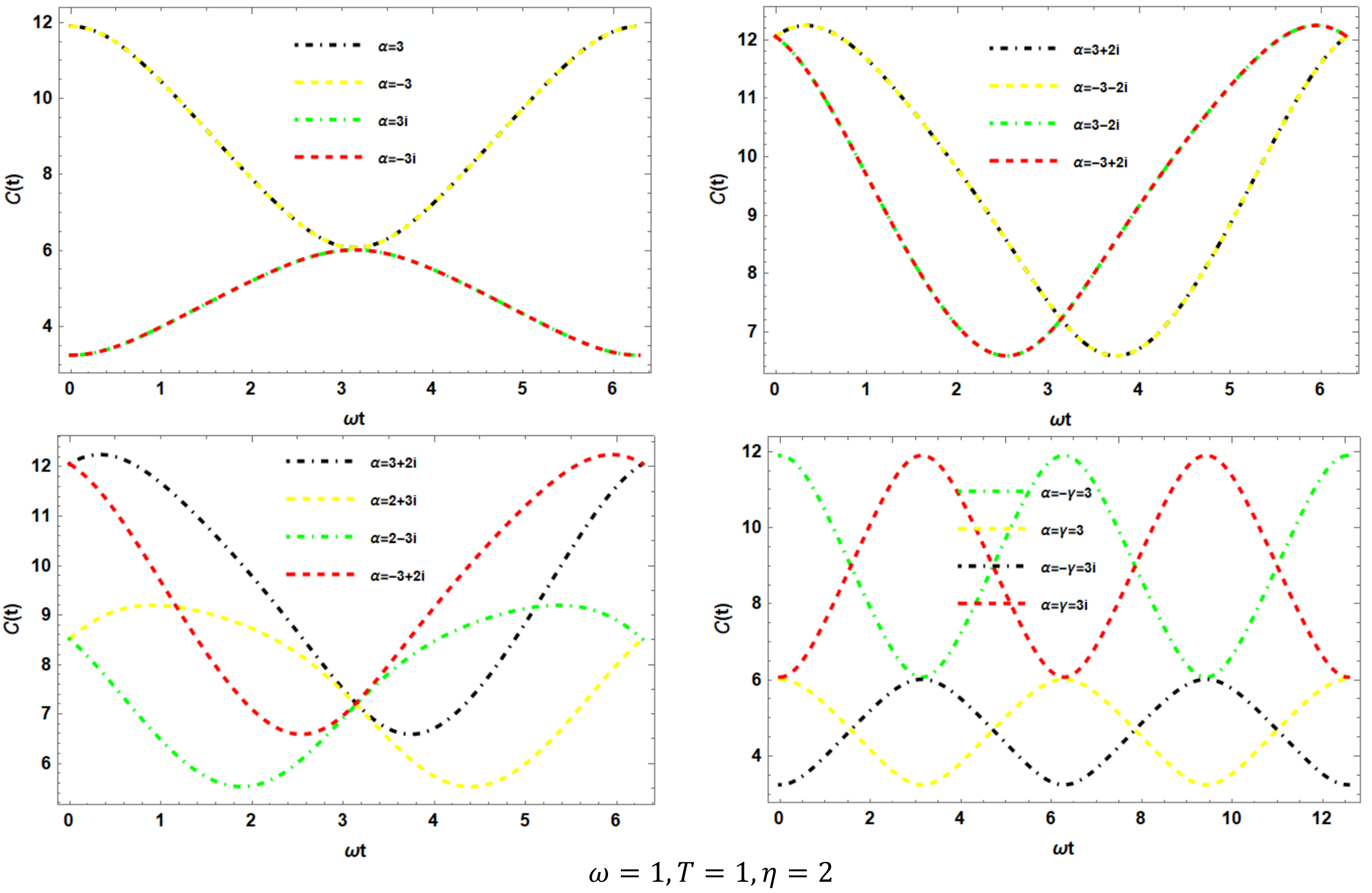}
\caption{{\it The complexity for two opposite excitations with various eigenvalues $\alpha$. In the last panel, we also compare the results with the complexity for two equal excitations (represented by the red and the orange lines).  }}
\label{twoop}
\end{figure}

Next, we turn our attention to the complexity for two opposite excitations. Some interesting results are shown in Fig. \ref{twoop}. From the figure, we conclude as follows.
\begin{enumerate}
\item Similar to the two equal excitation case, for general $\alpha$, there exists a relation\\ $\c_{(\alpha\,;-\alpha)}(t)=\c_{(-\alpha\,;\alpha)}(t)$. Moreover, the complexity for two states with conjugated eigenvalues are symmetric about the axis $t=\pi/\omega$ in the first period and hence $\c_{(\alpha^*\,;-\alpha^*)}(t)=\c_{(\alpha\,;-\alpha)}(2\pi/\omega-t)$.

\item Compared to the two equal excitation case, the two opposite excitation states will have a higher peak value of complexity if they have a larger absolute value of the real parts of the eigenvalues $|\Re\alpha|$, instead of the imaginary parts, see the lower left panel. In other words, the real parts of the eigenvalues $\alpha$ or $\gamma$ give greater influence in preparing two opposite excited states. This is very different from the two equal excitation case, which is more affected by the imaginary parts of the eigenvalues. In particular, from the lower right panel, we find that when $\alpha$ is real, $\c^{min}_{(\alpha\,;-\alpha)}=\c^{max}_{(i\alpha\,;-i\alpha)}$, implying that $\c_{(\alpha\,;-\alpha)}(t)\geq \c_{(i\alpha\,;-i\alpha)}(t')$. This strongly supports the above argument.

\item Furthermore, the last panel also tells us that when the eigenvalues are real or pure imaginary, the complexity for a two opposite excitation is related to that of a two equal excitation. For example, if $\alpha$ is real, then $\c_{(\alpha\,;-\alpha)}(t)=\c_{(i\alpha\,;i\alpha)}(t+\pi/\omega)$ and $\c_{(i\alpha\,;-i\alpha)}(t)=\c_{(\alpha\,;\alpha)}(t+\pi/\omega)$. The results, together with the relation $\mathcal{C}_{(i\alpha\,;i\alpha)}^{min}=\mathcal{C}_{(\alpha\,;\alpha)}^{max}$ for the two equal excitation case, lead to
   $\mathcal{C}_{(\alpha\,;-\alpha)}^{min}=\mathcal{C}_{(i\alpha\,;i\alpha)}^{min}
   =\mathcal{C}_{(\alpha\,;\alpha)}^{max}=\mathcal{C}_{(i\alpha\,;-i\alpha)}^{max}$.
    It implies that $\mathcal{C}_b(t)\leq \mathcal{C}_c(t')$ (or $\mathcal{C}_b{(t)}\geq \mathcal{C}_c{(t')}$) when only the real (or imaginary) parts of the eigenvalues are excited (here the subscript for the complexity follows sec.\ref{dvatrs}).


\end{enumerate}

\textbf{c) Two conjugated excitations ($\alpha=\gamma^\ast$).}

\begin{figure}[ht]
\centering
\includegraphics[width=450pt]{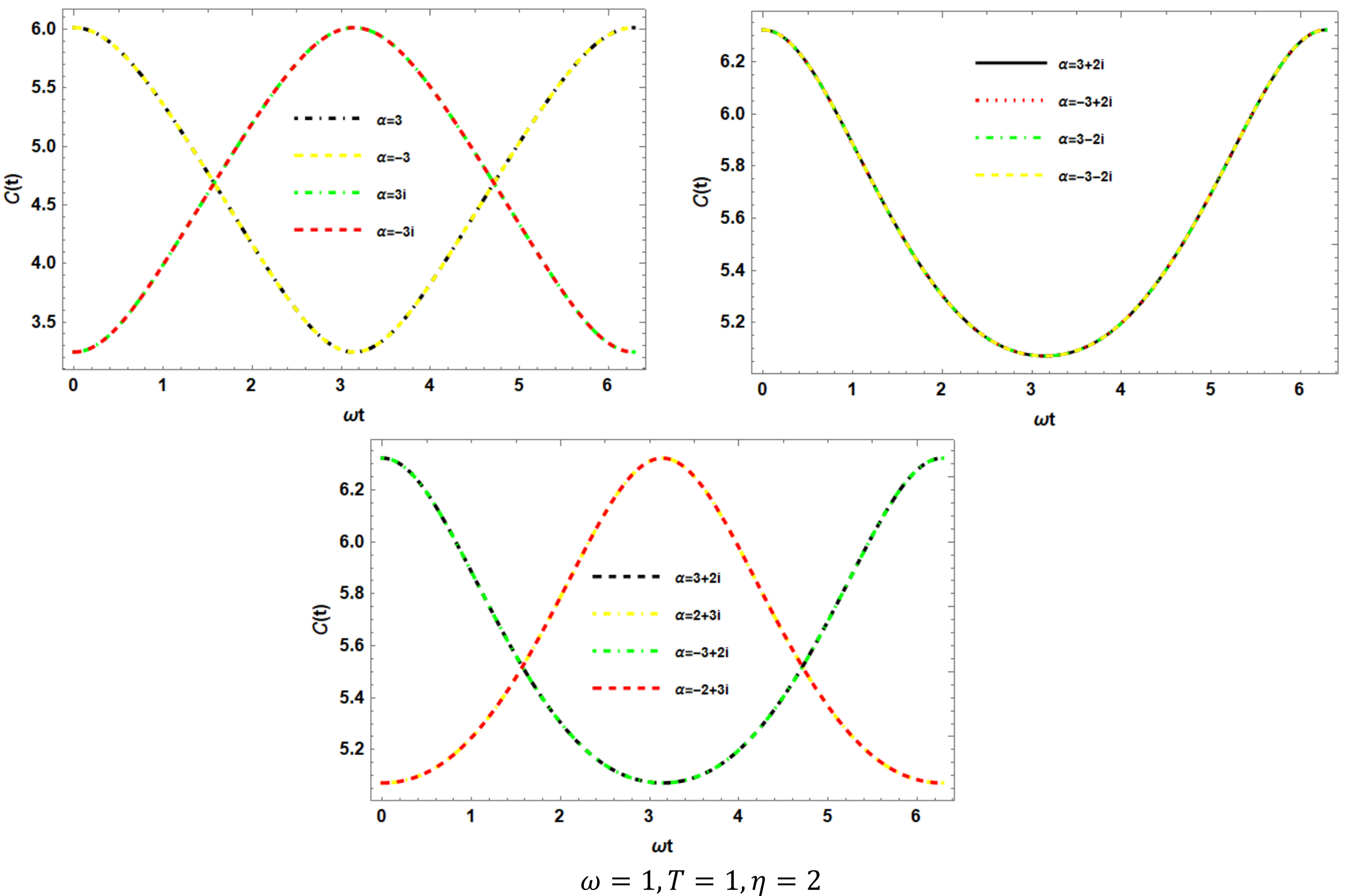}
\caption{{\it The complexity for two conjugated excitations with various eigenvalues $\alpha$.  }}
\label{twoco}
\end{figure}

At last, we close this part with the complexity for two conjugated excited states. Some numerical results are shown in Fig. \ref{twoco}. We conclude as follows.
\begin{enumerate}

\item First, the complexity for two conjugated excitations again respects the relation $\c_{(\alpha\,;\alpha^*)}(t)=\c_{(-\alpha\,;-\alpha^*)}(t)$. This is similar to the previous cases. Moreover, for general $\alpha$, the complexity is always symmetric about the axis $t=\pi/\omega$, leading to $\c_{(\alpha\,;\alpha^*)}(t)=\c_{(\alpha\,;\alpha^*)}(2\pi/\omega-t)$.

\item  From the upper right panel of Fig. \ref{twoco}, we find that unlike the previous cases, the complexity enjoys an enhanced symmetry $\c_{(\alpha\,;\alpha^*)}(t)=\c_{(\alpha^*\,;\alpha)}(t)=\c_{(\alpha^*\,;\alpha)}(2\pi/\omega-t)$. However, the first equality is simply a remanent of the fact that the states under consideration are invariant under the transformation $(\alpha\,,\gamma)\rightarrow (\gamma\,,\alpha)$, namely $|\psi_{CT}(\alpha\,,\gamma\,;t)\rangle=|\psi_{CT}(\gamma\,,\alpha\,;t)\rangle$. Thus, in general, one has $\c_{(\alpha\,;\gamma)}(t)=\c_{(\gamma\,;\alpha)}(t)$.

\item The lower panel tells us that the maximal value of the complexity for two conjugated excited states will be the same if the eigenvalues of the states simply exchange the absolute value of the real and the imaginary parts of the eigenvalues, namely $\c^{max}_{(\alpha\,;\alpha^*)}=\c^{max}_{(-\alpha\,;-\alpha^*)}=\c^{max}_{(\pm i\alpha\,;\mp i\alpha^*)}$. Compared with the aforementioned two types of excited states, this feature is unique. It implies that the maximal cost to prepare a two conjugated excited state from the reference state is insensitive to which part of the eigenvalue $\alpha$ is bigger or smaller as long as their absolute values $|\Re\alpha|\,,|\Im\alpha|$ are given.

\item Since the two conjugated excitations reduce to the two equal (opposite) excitations when only the real (imaginary) parts are excited, we again have the relation $\mathcal{C}_{(\alpha\,;-\alpha)}^{min}=\mathcal{C}_{(i\alpha\,;i\alpha)}^{min}
   =\mathcal{C}_{(\alpha\,;\alpha)}^{max}=\mathcal{C}_{(i\alpha\,;-i\alpha)}^{max}$, where $\alpha$ is real. It implies that
\bea
&\c_{c}(t)&\ge\c_{d}(t^\prime)=\c_{b}(t^\prime),\quad\quad
\eea
when only the real parts of the eigenvalues are excited and
\bea
\c_{b}(t)\ge\c_{d}(t^\prime)=\c_{c}(t^\prime),\quad\quad
\eea
 when only the imaginary parts are excited, where $t$ and $t^\prime$ are arbitrary. Compared to the analogous relations in section \ref{dvatrs}, one can immediately find that they are quite different. It tells us that different choices of the reference state not only affects the complexity for a single target state but also changes the relations between the complexities for different target states.

\end{enumerate}

\subsubsection{Comments on general case}
One may have noticed that the above results for the several special cases show some common features. This strongly motivates us to further study the complexity for the CT states with general eigenvalues. By scanning the parameters space, we find that there indeed exist some universal relations for the complexity. Without presenting more numerical results, we summarize the relations as follows and try to explain them from a physical or mathematical point of view.
\begin{enumerate}

\item The first relation is
\be \c_{(\alpha\,;\gamma)}(t)=\c_{(\gamma\,;\alpha)}(t)\,.\ee

It is simply a remanent of the fact that interchanging the eigenvalues between the two sides of the system does not change the state under consideration, namely $|\psi_{CT}(\alpha\,,\gamma\,;t)\rangle=|\psi_{CT}(\gamma\,,\alpha\,;t)\rangle$.

\item The second is the inversion relation
\be \c_{(\alpha\,;\gamma)}(t)=\c_{(-\alpha\,;-\gamma)}(t) \,.\label{cprelation2}\ee
It strongly implies that the complexity generally depends only on the various quadratic polynomials of the real and imaginary parts of the eigenvalues $\alpha$ and $\gamma$. This can be easily understood from our formula (\ref{cpgene}). The complexity for the ground state $\c_{(0)}$ does not depend on the eigenvalues. On the other hand, according to (\ref{CTSlambda}) and (\ref{vectornu}), one finds that the inversion $(\alpha\,,\gamma)\rightarrow (-\alpha\,,-\gamma)$ leads to $\lambda\rightarrow -\lambda$ and hence $\nu\rightarrow -\nu$. However, since the complexity depends quadratically on the vector $\nu$, the result is clearly invariant.

\item The third relation we found is
\be \c_{(\alpha^*\,;\gamma^*)}(t)=\c_{(\alpha\,;\gamma)}(2\pi/\omega-t)=\c_{(\alpha\,;\gamma)}(-t)\,.\ee
However, it is hard to prove since the rotation generator $K$ is time dependent as well as the complexity for the ground state. Furthermore, we also find
\be \c_{(0)}(t)=\c_{(0)}(\pi/\omega-t)=\c_{(0)}(-t) \,.\ee
Again, we do not know how to prove it, without solving the generator $K$ analytically. Nevertheless, the above result, together with (\ref{cpgene}) tells us that under the transformation $(\alpha\,,\gamma\,,t)\rightarrow (\alpha^*\,,\gamma^*\,,-t)$, the elements of the vector $\nu$ should transform as $\nu_i\rightarrow \pm \nu_i$. To examine what it means, we introduce $\lambda_p=(\lambda_{p_L}\,,\lambda_{p_R})\,,\lambda_q=(\lambda_{q_L}\,,\lambda_{q_R})$ and
 \be(M-I)^{-1}K\equiv
\left(
\begin{array}{cc}
  \hat{K}_{p_+} & -\hat{K}_{q_-} \\
  \hat{K}_{p_-} & -\hat{K}_{q_+}
\end{array}
\right)\,,\ee
where $\hat{K}_{p_\pm}\,,\hat{K}_{q_\pm}$ are $2\times 2$ partitioned matrixes. It follows that
\be \nu^T=\left(
\begin{array}{c}
  \hat{K}_{p_+} \lambda_p+\hat{K}_{q_-}\lambda_q \\
  \hat{K}_{p_-} \lambda_p+ \hat{K}_{q_+}\lambda_q
\end{array}
\right)\,.\ee

On the other hand, under the above transformation, we find that $\lambda_p\rightarrow \lambda_p$ and $\lambda_q\rightarrow -\lambda_q$. Thus, symmetry considerations indicate that under the time reversal, the partitioned matrixes $\hat{K}_{p_\pm}$ probably transform in an opposite way as $\hat{K}_{q_\mp}$, namely $(\hat{K}_{p_\pm}\,,\hat{K}_{q_\mp})\rightarrow (\epsilon\hat{K}_{p_\pm}\,,-\epsilon\hat{K}_{q_\mp})$, where $\epsilon=\pm 1$.

To check whether this is the case, we first consider the simplest case: the reference state is the Dirac vacuum. Using the results in sec.\ref{dvatrs}, we deduce
\bea
&& \hat{K}_{p_+}= \fft{\theta}{2}\left(\begin{array}{cc}
   \coth{(\ft\theta2)} & -\cos{\omega t} \\
  -\cos{\omega t} & \coth{(\ft\theta2)}
\end{array}\right) \,,\quad
\hat{K}_{q_-}= \fft{\theta}{2m\omega}\left(\begin{array}{cc}
  0 & -\sin\omega t \\
  -\sin\omega t & 0
\end{array}\right)  \,,      \nn\\
&& \hat{K}_{p_-}= \fft{m\omega\theta}{2}\left(\begin{array}{cc}
  0 & \sin\omega t \\
  \sin\omega t & 0
\end{array}\right) \,,\quad
\hat{K}_{q_+}= -\fft{\theta}{2}\left(\begin{array}{cc}
   \coth{(\ft\theta2)} & \cos{\omega t} \\
  \cos{\omega t} & \coth{(\ft\theta2)}
\end{array}\right) \,.
\eea
It is easy to see that under the time reversal, $(\hat{K}_{p_\pm}\,,\hat{K}_{q_\pm})\rightarrow \pm(\hat{K}_{p_\pm}\,,\hat{K}_{q_\pm})$. This gives us strong confidence that the above argument is reasonable. Moreover, for general Gaussian reference states, we can numerically verify that the partitioned matrixes exactly transform in the same way as the Dirac vacuum case.
\end{enumerate}

\section{Complexity for quantum field theory}\label{secqft}

In this section, we move to the circuit complexity for generalised coherent states in a $(1+1)$-dimensional free scalar field theory living on a cylinder with circumference $\math{L}$. Based on our experience for the two modes case, the numerical calculations for 2N modes will be straightforward but costly after we built the model of circuit complexity for the generalised coherent states in QFT. Hence, in this section, we would like to focus on analytical treatment of complexity in QFT. Along this line, we start with the Hamilton of the theory,
\ba\begin{aligned}\label{H1}
H=\int_{-\math{L}/2}^{\math{L}/2}dx\left(\frac{1}{2}\p(x)^2+\frac{1}{2}m^2\f(x)^2+\frac{1}{2}(\pd_x\f(x))^2\right)\,.
\end{aligned}\ea
We will regulate the field theory by a lattice model in which the lattice spacing is defined by
\ba\begin{aligned}
\d=\math{L}/N\,,
\end{aligned}\ea
where $N$ is the site number of the lattice arranged on the spatial circle. For simplicity we redefine the canonical variables as
\ba\begin{aligned}
Q_a=\f(x_a)\d\,,\ \ \ P_a=\p(x_a)\d\,,
\end{aligned}\ea
and impose periodic boundary conditions, $Q_{N+1}:=Q_1$ and $P_{N+1}:= P_1$. Then the Hamilton can be rewritten as
\ba\begin{aligned}
H=\sum_{a=1}^{N}\left(\frac{\d}{2}P_a^2+\frac{m^2}{2\d}Q_a^2+\frac{1}{2\d^3}\left(Q_a-Q_{a+1}\right)^2\right)\,.
\end{aligned}\ea
With the help of the Fourier transformation
\ba\begin{aligned}\label{QPk}
\tilde{Q}_k=\frac{1}{\sqrt{N}}\sum_{a=1}^{N}e^{2\p i k a/N}Q_a\,,\quad\quad\quad \tilde{P}_k=\frac{1}{\sqrt{N}}\sum_{a=1}^{N}e^{2\p i k a/N}P_a\,,
\end{aligned}
\ea
the Hamilton can be recast into a more compact form:
\ba\begin{aligned}
H=\sum^{N-1}_{k=0}\left(\frac{\d}{2}|\tilde{P}_k|^2+\frac{\w_k^2}{2\d}|\tilde{Q}_k|^2\right)\,,
\end{aligned}\ea
where the frequency is given by
\ba\begin{aligned}
\w_k=\sqrt{m^2+\frac{4}{\d}\sin^2\big(\frac{\p k}{N}\big)}\,,
\end{aligned}\ea
The canonical commutation relations are given by
\ba\begin{aligned}
\left[\tilde{Q}_k,\tilde{P}_{k'}\right]=i \d_{kk'}\,,\quad\quad \left[\tilde{Q}_k^\dag,\tilde{P}_{k'}^\dag\right]=i \d_{kk'}\,.
\end{aligned}\ea
In  view of the symmetry $\tilde{Q}_{N-k}^\dag=\tilde{Q}_{k}$, $\tilde{P}_{N-k}^\dag=\tilde{P}_{k}$ and $\omega_{N-k}=\omega_k$, we would like to introduce two new canonical quantities
\ba q_k=\left\{\begin{aligned}
\mathrm{Re} (\tilde{Q}_k)\,,\quad\quad k<\left[N/2\right]\\
\mathrm{Im} (\tilde{Q}_k)\,,\quad\quad k\geq\left[N/2\right]
\end{aligned}\right., \quad\quad p_k=\left\{\begin{aligned}
\mathrm{Re} (\tilde{P}_k)\,,\quad\quad k<\left[N/2\right]\\
\mathrm{Im} (\tilde{P}_k)\,,\quad\quad k\geq\left[N/2\right]
\end{aligned}\right.\,,
\ea
which obey the canonical commutation relations
\ba\begin{aligned}\label{newvari}
\left[q_k, p_{k'}\right]=i\d_{kk'}\,.
\end{aligned}\ea
Furthermore, in terms of these new canonical variables,  the Hamilton simplifies to
\ba\begin{aligned}
H=\sum^{N-1}_{k=0}\left(\d p_k^2+\frac{\w_k^2}{\d}q_k^2\right)\,.
\end{aligned}\ea
Therefore, the system can be viewed as a sum of $N$ independent harmonic oscillators with equal mass $m=1/2\d$ and frequency $\w_k$.

The whole system we are interested in is a double copy of  the free scalar theory. It is characterized by the Hamilton $H_L\otimes H_R$, as well as the canonical variables $\{\f_L(x),\f_R(x),\p_L(x), \p_R(x)\}$ or equivalently $\{ q_k^L,q_k^R,p_k^L,p_k^R \}$. We denote
\ba\begin{aligned}
\x^a \equiv\bigcup_{k=0}^{N-1}\{q_k^L,q_k^R,p_k^L,p_k^R\}\,.
\end{aligned}\ea
The annihilation and creation operators can be defined as
\ba\begin{aligned}
a^{L/R}_k&=\fft{1}{\sqrt{2m\omega_k}}\left(m\omega_k q_k^{L/R}+ip_k^{L/R}\right)\,,\\
a^{L/R\dag}_k&=\fft{1}{\sqrt{2m\omega_k}}\left(m\omega_k q_k^{L/R}-ip_k^{L/R}\right)\,.
\end{aligned}\ea
They obey the commutation relations
\be [a_k^L\,,a_{k'}^{L\dag}]=[a_k^R\,,a_{k'}^{R\dag}]=\delta_{kk'} \,.\label{commutationkk}\ee
 As expected, for a free field theory, the modes with different ``momenta" $k$ simply commute with each other. From this and the results for a single harmonic oscillator, it will be straightforward to construct some states of interest for the total system. For example, the TFD state can still be defined as
 \be |\mathrm{TFD}\rangle=U(\beta)|0\rangle_L|0\rangle_R  \,,\ee
 where the anti-Hermitian operator $U(\beta)$ now is given by
\ba\begin{aligned}
U(\b)=\prod_{k=0}^{N-1}U_k(\beta)=\exp\left[\sum_{k=0}^{N-1}\q_k(\b)(a_k^{L\dag}a_k^{R\dag}-a_k^La_k^R)\right]\,,
\end{aligned}\ea
where $\q_k(\b)=\text{arctanh}\big(e^{-\b\w_k/2}\big)$. It turns out that the TFD state can be written as a tensor product as
\be |\mathrm{TFD} \rangle=\bigotimes_{k=0}^{N-1}|\mathrm{TFD}\rangle_k  \,,\ee
where $|\mathrm{TFD}\rangle_k$ stands for the TFD state for a single harmonic oscillator. Likewise, the coherent thermal (CT) state can still be defined by (\ref{CTS1}). One finds
\be |\mathrm{CT} \rangle=\prod_{k=0}^{N-1}\mathrm{exp}\big[ \alpha_k a_k^{L\dag}+\gamma_k a_k^{R\dag}-\alpha_k^* a_k^L -\gamma_k^* a_k^R \big]|\mathrm{TFD}\rangle= \bigotimes_{k=0}^{N-1}|\mathrm{CT}(\alpha_k\,,\gamma_k) \rangle\,.\ee
Of course, the relation can be generalised to the time dependent states directly
\ba\begin{aligned}
|\Y_{CT}(t)\rangle=\bigotimes_{k=0}^{N-1}|\psi_\text{CT}(\a_k,\g_k, t)\rangle\,,
\end{aligned}\ea
where $|\psi_\text{CT}(\a_k,\g_k, t)\rangle$ describes the same state (time dependent CT state) defined in Eq. (\ref{ctstime}) for a single harmonic oscillator except that now $m\rightarrow 1/2\delta\,,\omega\rightarrow \omega_k$.

By definition, it is easy to see that the covariance matrix for such tensor product states can be recast into a corresponding tensor plus form. For the CT state, we have
\ba\begin{aligned}
G_T(|\Y_{CT}(t)\rangle)=\bigoplus_{k=0}^{N-1} G_T(|\psi_\text{CT}(\a_k,\g_k, t)\rangle)\,,
\end{aligned}\ea
where $G_T(|\psi_\text{CT}(\a_k,\g_k, t)\rangle)$ has the same expression as Eq. (\ref{comatrix}) except that the parameters  $(\a,\g,\w,\q)$ now should be replaced by $(\a_k,\g_k,\w_k,\q_k)$.

To proceed, we would like to choose the Dirac vacua as the reference state, i.e., the ground state for the Hamilton \eq{H1}. Its covariance matrix is given by
\ba\begin{aligned}
G_R=\bigoplus_{k=0}^{N-1} G_R^k\,,
\end{aligned}\ea
where
\ba\begin{aligned}
G_R^k=\left(
\begin{array}{cccc}
 \frac{\d}{\w_k} & 0 & 0 & 0 \\
 0 & \frac{\d}{\w_k} & 0 & 0 \\
 0 & 0 & \frac{\w_k}{4\d} & 0 \\
 0 & 0 & 0 & \frac{\w_k}{4\d}
 \end{array}\right)\,.
\end{aligned}\ea
Following closely the discussions in sec.\ref{dvatrs}, we deduce the complexity
\be\label{cpctsfield} \mathcal{C}=\sum_{k=0}^{N-1}\theta_k\, \text{csch}\big(\ft{\theta_k}{2} \big) \sqrt{\big(|\alpha_k|^2+|\gamma_k|^2+2 \big)\cosh\theta_k+2(\Im\alpha_k \Im\gamma_k-\Re\alpha_k \Re\gamma_k)\sinh\theta_k-2} \,.\ee
The result is simply a sum of the complexity for the N-independent harmonic oscillators
\be \mathcal{C}=\sum_{k=0}^{N-1} \mathcal{C}_k(\alpha_k\,,\gamma_k\,,\omega_k) \,.\ee
It is not hard to believe that the relation is valid to more general reference states.

Alternatively, we may take the reference state to be the ground state of the ultralocal Hamilton, which is given by
\ba\begin{aligned}\label{H2}
H_\text{ul}=\int_{-\math{L}/2}^{\math{L}/2}dx\left(\frac{1}{2}\p(x)^2+\frac{1}{2}\m^2\f(x)^2\right)\,,
\end{aligned}\ea
where $\m$ is a constant parameter playing the role of the mass. Using the canonical variables $(q_k\,,p_k)$, we find
\ba\begin{aligned}
H_\text{ul}=\sum^{N-1}_{k=0}\left(\d p_k^2+\frac{\m^2}{\d}q_k^2\right)\,.
\end{aligned}\ea
Notice that the frequency is a same constant $\omega=2\mu$ for all the modes. The covariance matrix for this ground state turns out to be
\ba\begin{aligned}
G_R=\bigoplus_{k=0}^{N-1} G_R^k\,,
\end{aligned}\ea
with
\ba\begin{aligned}
G_R^k=\left(
\begin{array}{cccc}
 \frac{\d}{2\m} & 0 & 0 & 0 \\
 0 & \frac{\d}{2\m} & 0 & 0 \\
 0 & 0 & \frac{\m}{2\d} & 0 \\
 0 & 0 & 0 & \frac{\m}{2\d}
 \end{array}\right)\,.
\end{aligned}\ea
Notice that $G_R^k$ can be viewed as the covariance matrix (\ref{generalCM}) for a general Gaussian reference state with the parameter $\eta_k=\mu/\omega_k$. This implies that the complexity corresponding to this reference state can be obtained by
\ba\begin{aligned}
\math{C}=\sum_{k=0}^{N-1}\math{C}_k(\a_k,\g_k,\w_k,\h_k)\,.
\end{aligned}\ea
Again, the result is simply a sum of the complexity $\math{C}_k(\a_k,\g_k,\w_k,\h_k)$ for the N single harmonic oscillators.

\section{Conclusions}

In the inspiring paper \cite{Chapman:2018hou}, the authors first calculated the circuit complexity for time-dependent TFD states using the covariance matrix approach. In the present paper, we extended their analysis to consider the complexity of the generalised coherent states in thermal field dynamics. In our case, the target state is not Gaussian anymore and has a nonvanishing one-point function.

We started from the harmonic oscillator system.
Based on the construction of the Glauber coherent state and the thermal vacuum state, we introduced Coherent Thermal (CT) state and Thermal Coherent (TC) state, respectively. Also, we found they are related through Eq. (\ref{connection}), and the time-dependent TC state is related to the corresponding time-dependent CT state by Eq. (\ref{connection2}). Thus we found the complexity for a TC state has a simple connection with the one for the corresponding CT state, as shown in Eq. (\ref{cp2}).

As the one-point function of the target state is not vanishing,  the covariance matrix approach used in \cite{Chapman:2018hou} cannot be directly applied in evaluating the circuit complexity for the CT state, as the symmetric two-point function is not enough to determine a CT state. Nevertheless, by examining the properties of the CT state carefully, we developed the generalized covariance matrix approach  and applied it to compute the circuit complexity.  We first introduced the one-point function in Eq. (\ref{onepo}), and from the transformation law of the 1-point and two-point functions we define  the generalized covariance matrix  Eq. (\ref{excm}). The corresponding generators preparing our non-Gaussian state form a group structure $\mathbb{R}^{2N} \rtimes \mathrm{Sp}(2N\,,\mathbb{R})$. The essential point is that the optimal geodesic is still be generated by the horizontal generator, if the reference state is still Gaussian.  After some careful analysis, we derived the circuit complexity for a CT state in Eq. (\ref{cplex2}). This expression is one of the most important results in this work, so here we write it again,
\ba
\math{C}^2=\text{Tr}\big (K^2 \big)+\n g_R\n^T\,. \label{CTcircuitcomplexity}
\ea
The notable feature of this formula is that there is an extra piece $\n g_R\n^T$ contributing to the complexity.

With the formula \eqref{CTcircuitcomplexity}, we were allowed to study the circuit complexity of the CT state. We  first chose the Dirac vacuum as the reference state, and obtained the complexity which is given in Eq. (\ref{cplex2}), Based on the formula \eqref{CTcircuitcomplexity}. Surprisingly, we found this result is independent of time. With this formula in hand, we examined some special cases including one single excitation, two equal excitations, two opposite excitations and two conjugated excitations. From these results, we discover that the complexity  with two conjugated excitations is always smaller than the one with two equal excitations or two opposite excitations fixing $\alpha$.  Interestingly, the complexity with two equal excitations is more sensitive to the real part of $\alpha$ while the complexity with two opposite excitations relies more on the imaginary part of $\alpha$. In addition, we also compared the influence of single excitation and two excitations with the same $\alpha$, we found which complexity in two cases is larger not only depends on the temperature, but also relies on the value of $\alpha$. This fact implies, the target state with two excitations is highly entangled, and the complexity also reveals the characteristics of entanglement qualitatively.

Furthermore we considered the case that the reference state is a more general Gaussian state instead of  the Dirac vacuum. The calculation of the circuit complexity is straightforward, but the details were much more complicated and were asked for numerical method. Similarly, we looked at four different cases with the single excitation and double excitations including the equal, the opposite and the conjugated excitations. For each case, we found the function of the complexity is time-dependent with a period being $2\pi/\omega$ rather than $\pi/\omega$. This is due to the  extra piece in \eqref{CTcircuitcomplexity}. We presented some numerical results for the four different cases and read some interesting findings in sec. \ref{gaussion}. In particular  we investigated how various parameters affect the complexity. These parameters include $\eta$ which reflects the initial state,  the temperature $T$ and the level of the excitations $P$ or $Q$ . We studied the symmetry and translation of the complexity in time. Moreover, we gave a study on the exchange symmetry and the parity symmetry of the real part and imaginary part of $\alpha=P+iQ$. Similar to the discussion on the Dirac vacuum, we  compared the complexities of different kinds of excitations with fixed $\alpha$, especially the one of double excitations. For example, we found that to some extent the imaginary part of $\alpha$ or $\gamma$ gave more significant influence in preparing two equal excited states with the same $|\alpha|$ or $|\gamma|$ from the reference state,  while the real parts affected more significantly on the two opposite excitations. For two conjugated excitations, there is no obvious dependence on the real parts or imaginary parts of $\alpha$ or $\gamma$. 

In sec. \ref{secqft}, we briefly showed that  the previous analysis can be extended to a free scalar field theory taking  the Dirac vacuum  as the reference state. Our study was preliminary and could be extended to different directions.
We would like to study the circuit complexity in QFT taking a general Gaussian state as the reference state in the future work since there are too many parameters for the generalised coherent states. Besides, in the light of the first law of the complexity proposed in \cite{firstlaw} and revisited in \cite{firstlaw1}, it is also interesting to verify whether the first law is valid when we extent to the  generalised coherent states. Another interesting topic is to generalize our study to the fermionic case. 

\section*{Acknowledgments}
Z.Y. Fan was supported in part by the National Natural Science Foundations of China with Grant No. 11805041, No. 11873025 and No. 11575270. B. Chen was in part supported by NSFC Grant  No.~11335012, No.~11325522 and No. 11735001. M. Guo is supported by NSFC Grant No. 11947210. And he is also funded by China National Postdoctoral Innovation Program 2019M660278. J. Jiang is supported by National Natural Science Foundation of China (NSFC) with Grants No. 11775022 and No. 11873044.

\end{document}